\documentclass[sigconf]{acmart}

\usepackage{booktabs} 
\setcopyright{rightsretained}
\usepackage{algorithm}
\usepackage{caption}
\usepackage{subcaption}
\usepackage{algpseudocode}

\usepackage{comment}
\usepackage{epstopdf}
\usepackage{graphicx}
\usepackage{mwe}

\usepackage{graphics} 
\usepackage{amsmath}
\usepackage{pbox}
\usepackage{flushend}
 \usepackage{url}

\DeclareCaptionLabelFormat{mylabel}{#1 #2.\hspace{1.5ex}}
\graphicspath{{./figures/}}
\def \sys {\textit{HybridLoc}}

\DeclareMathOperator*{\argmax}{arg\,max}
\begin{document}
\title{The Tale of Two Localization Technologies: Enabling Accurate Low-Overhead WiFi-based Localization for Low-end Phones}

\author{Ahmed Shokry$^{\dag}$, Moustafa Elhamshary$^{\dag\ddag}$, Moustafa Youssef$^{\dag}$
}
\affiliation{%
  \institution{$\dag$ Wireless Research Center, Egypt-Japan University of Science and Technology (E-JUST), Alexandria, Egypt.}
  \institution{$\ddag$ Faculty of Engineering, Tanta University, Egypt.}
}

\begin{abstract}
WiFi fingerprinting  is one of the mainstream technologies for indoor
localization. However, it requires  an initial  calibration
phase during which the fingerprint database is built manually by site surveyors. This process is  labour intensive, tedious, and  needs to be repeated with any change in  the environment. 
 While a number
of recent systems have been introduced to reduce the calibration effort through RF propagation models  and/or crowdsourcing, these still 

 have some limitations.  Other approaches use the recently developed iBeacon technology   as an alternative to WiFi for indoor localization.
However, these beacon-based solutions are  limited to a small subset of high-end phones.

In this paper, we  present \sys{}:  an accurate  low-overhead indoor localization system. The basic idea \sys{} builds on is to leverage the sensors of high-end phones to enable localization of lower-end phones. Specifically, the WiFi fingerprint is crowd-sourced by opportunistically collecting WiFi-scans labeled with location data obtained from BLE-enabled high-end smart phones. These scans are used to automatically construct the WiFi-fingerprint, that is used later to localize any lower-end cell phone with the ubiquitous WiFi technology. \sys{} also has provisions for handling the inherent error in the estimated BLE locations used in constructing the fingerprint as well as to handle practical deployment issues including the noisy wireless environment,  heterogeneous devices, among others.

Evaluation of \sys{}  using Android phones shows that it can provide accurate localization in the same range as manual fingerprinting techniques under the same deployment conditions. Moreover, the localization accuracy on low-end phones supporting only WiFi is comparable to that achieved with high-end phones supporting BLE.  This accuracy is achieved with no training overhead, is robust to the different user devices, and is consistent under environment changes.

\end{abstract}

\begin{CCSXML}
<ccs2012>
 <concept>
  <concept_id>10010520.10010553.10010562</concept_id>
  <concept_desc>Computer systems organization~Embedded systems</concept_desc>
  <concept_significance>500</concept_significance>
 </concept>
 <concept>
  <concept_id>10010520.10010575.10010755</concept_id>
  <concept_desc>Computer systems organization~Redundancy</concept_desc>
  <concept_significance>300</concept_significance>
 </concept>
 <concept>
  <concept_id>10010520.10010553.10010554</concept_id>
  <concept_desc>Computer systems organization~Robotics</concept_desc>
  <concept_significance>100</concept_significance>
 </concept>
 <concept>
  <concept_id>10003033.10003083.10003095</concept_id>
  <concept_desc>Networks~Network reliability</concept_desc>
  <concept_significance>100</concept_significance>
 </concept>
</ccs2012>  
\end{CCSXML}

\ccsdesc[300]{Human-centered computing~Ubiquitous and mobile computing}

\keywords{Indoor Localization;  WiFi Fingerprinting; BLE-based Localization}
\maketitle
\section{Introduction}

Recent years have witnessed the advent of  indoor localization  systems  harnessing the many untapped capabilities of the smartphones \cite{youssef2015towards,elhamshary2017towards}.    
The development of wireless technology in smartphones and  the widespread availability of IEEE 802.11 have  enabled  the introduction of  many indoor localization solutions that can provide meter-level accuracy.  WiFi-based indoor localization techniques leverage   the Received Signal Strength (RSS)  overheard from WiFi access points as  the metric for the location determinations, building a WiFi fingerprint to combat the noisy wireless channel.  Typical fingerprint-based WiFi localization techniques work in two  phases: The first  phase is initial training  (i.e., calibration) during which RSS measurements (i.e., fingerprints) received from the multiple access points (APs) installed in the area of interest are recorded at known locations. Then, in the tracking phase, RSS measurements from  the overheard APs at an unknown location are matched against the  fingerprint database to determine the best location match either deterministically  or probabilistically.  Nonetheless,  the  deployment cost is prohibitive as  the WiFi calibration process    is time consuming, labour intensive and vulnerable to environmental dynamics. To tackle this problem, a number of approaches for automating the fingerprinting process  have been proposed including using RF  propagation models  \cite{eleryan2011synthetic,ji2006ariadne},  combining RF localization with other sensors \cite{wang2012no},  or   crowdsourcing  the fingerprint; where  users  perform the required survey process  in realtime either implicitly \cite{wang2012no,rai2012zee} or explicitly \cite{park2010growing}.  
These techniques, however, suffer from lower accuracy; require explicit user intervention; or work only on high-end phones.

To further address the issues of WiFi-based localization, 

 the iBeacon technology has been introduced. Beacons are cheap, portable, and energy-efficient devices based on the BlueTooth Low Energy (BLE) standard that can be installed at known locations in the area of interest. iBeacons periodically broadcast their  identifier  along with other location  information which  can be  overheard by nearby compatible  smartphones.  A number of commercial solutions leverage iBeacons to provide proximity-based localization \cite{locatify,estimote} and, more recently, researchers have started to use them to provide more accurate continuous user tracking \cite{elbakly2016robust,elbakly2016cone}. Nonetheless, the BLE technology is currently limited to high-end smart phones, limiting their ubiquitous deployment on typical commodity devices with the majority of the users. In addition, the opportunity of using the existing WiFi-infrastructure for localization is missed in this case. 

In this paper, we combine the best of both localization technologies. Specifically, we present \sys{}: an accurate,  robust, low-overhead, and ubiquitous \textbf{\textit{WiFi-based}} indoor localization system. \sys{} targets enabling WiFi-based localization for low-end phones in buildings with the iBeacon infrastructure while \textbf{\textit{removing the traditional WiFi calibration overhead}}.
The basic idea is to automatically crowdsource the construction of the WiFi fingerprint leveraging the high-end phones supporting the iBeacon technology. In particular, building users with \textit{high-end smart phones} will scan for \textbf{\textit{both}} BLE and WiFi APs concurrently in the area of interest while using the system. BLE scans are used to get an estimate of the user location. The user location is used with the scanned WiFi signals to construct the WiFi fingerprint in an automatic manner. This  WiFi fingerprint database grows incrementally by users with high-end phones visiting the  area of interest. Later,  users with low-end phones, i.e. those that support only WiFi but not BLE, can provide \sys{} with the WiFi scans to get an estimate of their location. Therefore, by leveraging the ground-truth location information ``donated'' by users with high-end phones, \sys{} can provide ubiquitous localization to any WiFi-enabled phones in areas with iBeacons deployment.

To achieve \sys{}'s goals, a new set of challenges still need to be addressed: First, the ground-truth location obtained from the BLE localization has an inherent error in the order of few meters. This error leads to  a mis-assignment of WiFi scans to the wrong fingerprint point. Second, to construct a probabilistic fingerprint for localization; which is proven to provide better accuracy than deterministic techniques \cite{youssef2003optimality,youssef2005horus,ibrahim2012cellsense}; traditionally a user has to stay  at each fingerprint point for a certain amount of time to construct the signal strength histogram. This adds significantly to the overhead of the fingerprint construction process, and cannot be performed while the users are moving naturally in the building.  Third, different phones will measure the RSS differently at the same location due to different WiFi chips, form factors, or chip placement inside the phone. This device heterogeneity needs to be addressed to avoid excessive localization error or per device-calibration. Finally,  due to the noisy wireless channel, there may be some  missing APs in successive scans that lead to a mismatch between the set of heard APs at the same location. \sys{} presents a number of modules to address  these challenges.  

We have implemented \sys{}  on different Android phones and  evaluated
 it in a university building instrumented with the iBeacon BLE technology and the already installed WiFi infrastructure. 
 Our results show that \sys{} can achieve a consistent median accuracy of 4.1m under different scenarios, which is comparable to the accuracy of manual fingerprinting techniques under the same conditions. In addition, the system accuracy on low-end phones supporting only WiFi is comparable to that achieved with high-end phones supporting BLE. This accuracy is achieved with no training
 overhead and is robust to the different user devices, and wireless channel noise.

The rest of the paper is organized as follows:  Section~\ref{sec:over} presents an overview on how \sys{} works and introduces the mathematical model.
Section~\ref{sec:sys}  gives the details of \sys{} and how it handles different practical considerations. We evaluate the system performance in Section \ref{sec:eval} and compare it to the state-of-the-art. Section~\ref{sec:related} discusses related work. Finally, Section~\ref{sec:conc} concludes the paper.

\section{Overview and Mathematical Model}
\label{sec:over}
In this section, we start by an overview  of how \sys{} works to illustrate the high-level flow of information through the system architecture. 
Then, we  discuss the mathematical model of our approach. 
\begin{figure}[!t]
\centering
\includegraphics[width=0.5\textwidth]{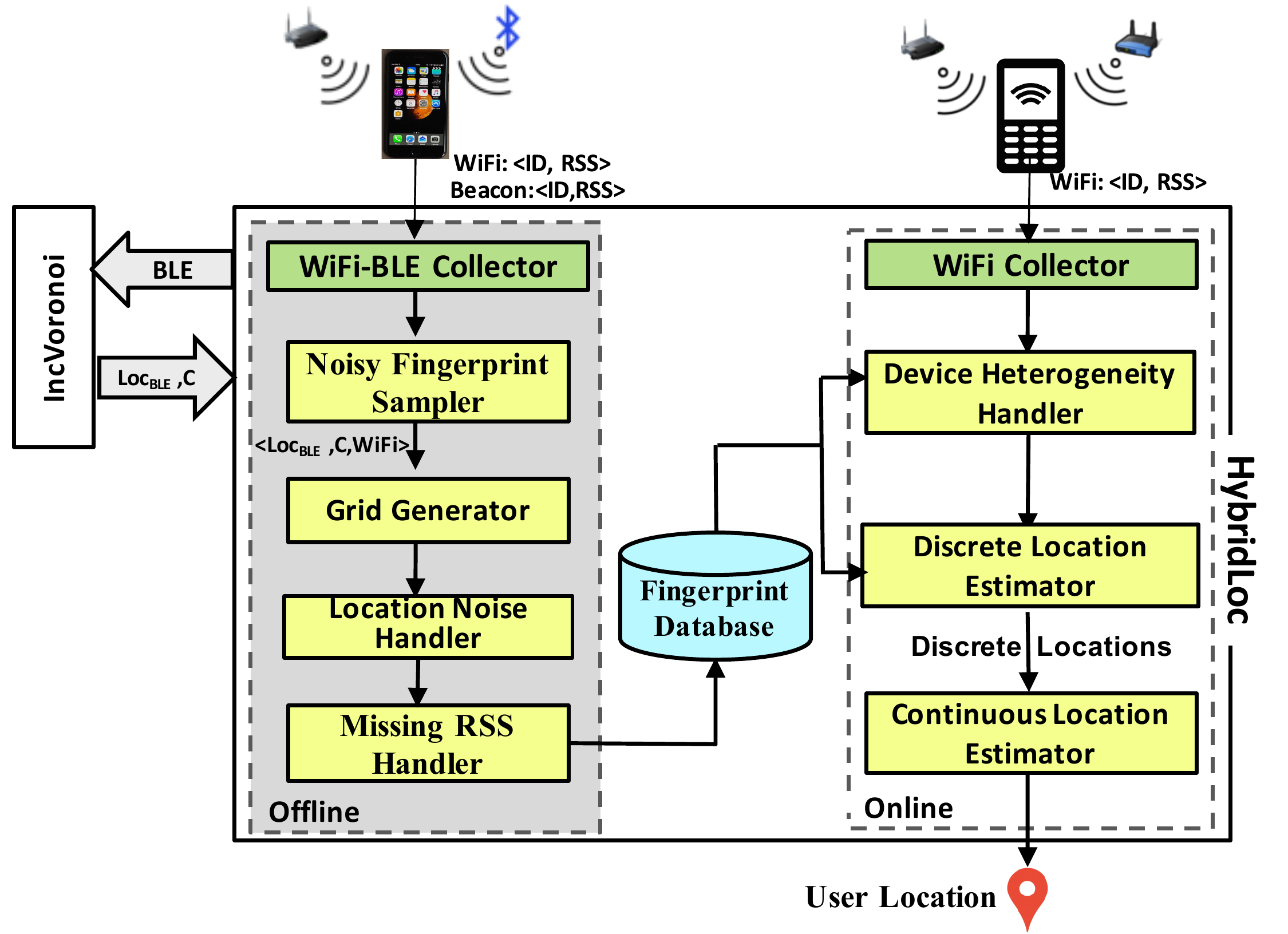}
\caption{\sys{} system architecture.}
 \label{fig:arch}
\end{figure}
\subsection{System Overview}
 Figure  \ref{fig:arch} shows \sys{} architecture.  
 \sys{} is designed to be deployed in areas where the iBeacon technology is already installed for localization and WiFi is deployed for coverage. Therefore, by default, it can provide BLE-based localization to users with high-end phones that support the iBeacon technology. \sys{} works in two phases: an offline fingerprint construction phase and an online tracking phase.
  During the offline phase, \sys{} aims to  \textit{\textbf{automatically}} build  a probabilistic  WiFi  fingerprint, where the RSS histogram for each AP at given locations in the area of interest is estimated.
  
 As users with high-end phones move naturally inside the building, the \sys{} software installed on their phone continuously scans the environment for \textit{\textbf{both}} BLE and  WiFi  signals transmitted from the  installed beacons and WiFi APs respectively.   These WiFi and BLE scans are submitted to the \textbf{Noisy Fingerprint  Sampler} module. This module forwards the collected vector of  BLE RSSs to a BLE-based indoor localization technology to get an estimate  of the user with high-end phones location\footnote{Note that these users are incetivized to install \sys{} software on their devices to navigate using the BLE technology.}. 
 \textbf{ Without loss of generality}, we use  IncrVoronoi  \cite{elbakly2016robust} as an example of a calibration-free BLE-based indoor localization system due to its high accuracy, robustness to heterogeneity in user devices, and adaption to dynamic changes in the environment and APs transmit power. In addition, IncVoronoi also provides a confidence measure of the estimated location \cite{elbakly2016cone}. 
  This location label is  annotated with  the collected WiFi scan at that  point  and the pair is recorded  a candidate WiFi fingerprint in the database.

To build the WiFi histogram at the every location, traditionally the user has to stay for a certain period of time in the same location, which may be inconvenient for users.  To reduce this overhead of the histogram construction  (i.e. allow building the histogram while the user is continuously moving), \sys{}  uses a  gridding approach, where the area of interest is divided into square cells of arbitrary size using  the \textbf{Grid Generator} module. The RSS histograms are then built using the collected fingerprint points inside each grid. This not only removes the extra overhead of fingerprint construction  but also increases the scalability of system as the fingerprint size can be arbitrarily reduced by increasing the cell size.
However, as the cell size increases, the accuracy and  computational-efficiency decrease. Thus,  the cell size  (density of the grid) should be configured by the system designer to trade-off accuracy and computational overhead.

Given that  the BLE-based  estimated locations have inherent noise,  which may results into many mis-assignment of WiFi scans to fingerprint points; to mitigate this effect;  the  \textbf{Location Noise Handler} module incorporates different approaches to handle  these noisy location labels building on the estimated location confidence provided by the IncVoronoi system.

The \textbf{Missing  RSS Handler} module further handles the cases where some RSSs are not heard in different scans from some APs during the training and/or tracking  phases, due to the wireless channel noise, possibly limiting  the localization accuracy.   

 The \textbf{Device Heterogeneity Handler} module provides a mathematical approach to make the system independent from the used device, hence increasing the system robustness to different heterogeneous phones.

 Finally, during the online phase, end users; even those with phones that support WiFi only but not BLE; are  tracked in realtime  by forwarding the scanned WiFi APs that can be heard at the current  unknown user location to the \sys{} WiFi tracking service. Specifically, the \textbf{Discrete Location Estimator} module consults the constructed WiFi probabilistic fingerprint and the center of the discrete grid that has the maximum probability is returned as the estimated user location. 
 To enable user tracking in the continuous space,  the \textbf{Continuous Location Estimator} module further  processes the  estimated discrete  grid locations to return a more accurate estimate of the user location. 

Note that since the system is crowdsensing-based, the fingerprint is continuously being updated in the background with the data collected from system users with high-end phones, allowing \sys{} to counter environment changes.

\subsection{Mathematical Model}
Without loss of generality, we assume a 2D area where $n$ WiFi APs and $m$ BLE beacons have been installed.  A user carrying a device at an unknown location $l$ scans for the nearby APs and beacons.
The scanned WiFi APs  can be represented as a vector  $s$ = $(s_1,..., s_q)$, where $q \le n$ is the number of heard APs and each $s_i$  is the signal strength of the $i^{th}$ heard AP. The scanned beacons  can be represented as a vector   $b$ = $(b_1,..., b_p)$,, where $p\le m $ is the number of heard beacons and each $p_i$  is the signal strength of the $i^{th}$ heard beacon. 

During the offline phase, where the fingerprint is constructed using high-end phones, the BLE scan ($b$) is forwarded to the IncVoronoi system to estimate the user location $l_b$ and its confidence $c$ (represented as a circle with radius $c$ around the estimated location $l_b$). A location-tagged WiFi scan $(l_b, c, s)$ is assigned to a specific grid cell $g$ out of the possible $\mathbb{G}$ grids in the area of interest. Each grid cell $g$ is represented by a single location $g_l$. 

During the online tracking phase, our problem is that, given some WiFi RSS vector $s$,  we want to find the grid cell $g\in \mathbb{G}$ that maximizes the probability $P(g|s)$. The estimated user location is $g_l$, the representative location of the cell with the highest probability.  

Table~\ref{notation} summarizes the notations used in the paper. 

\begin{table}[!t]
\centering
\begin{tabular}{|p{1.2cm}|l|}\hline
   Symbol & Description \\ \hline \hline
      $n$ &\pbox{10cm}   {Total  number of installed WiFi APs.}  \\ \hline
    $m$ & \pbox{10cm}   {Total number of installed BLE beacons.}  \\ \hline
 $s$ &\pbox{10cm}   {Vector of RSS from the heard APs in a scan.} \\ \hline
      $q$ &\pbox{10cm}    {Number of WiFi APs in a given scan ($q=|s|$).}\\ \hline
    $b$&\pbox{10cm}  {Vector of RSS from the heard beacons in a scan.}   \\ \hline
       $p$ &\pbox{10cm}   {Number of BLE beacons  in a given scan ($p=|b|$).}\ \\ \hline
       $l_b$& Estimated ground-truth location by IncVoronoi (BLE-based). \\ \hline
        $c$&  Degree of confidence in IncVoronoi estimated location.\\ \hline
       $ \mathbb{G}$  & Universe of grid cells in the area of interest.  \\ \hline
       $g$  &\pbox{10cm}{A specific grid cell,  $g \in \mathbb{G}$.}   \\\hline
        $g_l$  &\pbox{10cm}{Representative location of a grid cell.}   \\\hline
   
        $\mu_i$  &\pbox{10cm}{Average RSS from $AP_i$.}   \\\hline
        $\sigma_i$  &\pbox{10cm}{Standard deviation of RSS from $AP_i$.}   \\\hline

        $k$  &\pbox{10cm}{History window size used to estimate the continuous user \\location.}   \\\hline
        $ N_s$  &\pbox{10cm}{Number of  scans used in the training .}   \\\hline     
        $ G_S$  &\pbox{10cm}{The cell grid length (i.e., grid spacing). }   \\\hline     
         $ \lambda$  &\pbox{10cm}{The  common offset between RSSs collected by different\\ phones. }   \\\hline        
  \end{tabular}
  \caption{Notation table}
   \label{notation}
\end{table}
\section{The {HybridLoc} System}
\label{sec:sys}
In this section, we present the details of  the \sys{}  system architecture   for  indoor  localization. We  elaborate the following main functionalities: WiFi Fingerprint  construction (offline) phase  and tracking (online) phase. 
In addition,  we discuss  the different practical challenges that need to be addressed by \sys{} to enable its realtime deployment.

\subsection{Construction of the WiFi Fingerprint- Offline Phase}
This module aims to  automatically construct the WiFi fingerprint. It addresses a number of challenges including handling the error in the ground-truth BLE locations, missing APs, and heterogeneous devices. 

The input to this module is the WiFi and BLE scans collected by high-end smart phones. The module first queries the employed  zero-calibration BLE-based localization  (i.e., \textit{IncVoronoi} \cite{elbakly2016robust}) to retrieve and estimate of the user location given the BLE beacon RSS vector.   \textit{IncVoronoi} is based on the idea that the relative relation between the received signal strength from two APs at a certain location reflects the relative distance from this location to the respective APs. Building on this, it incrementally reduces the user ambiguity region based on refining the Voronoi tessellation of the area of interest. It also provides a confidence measure for the estimated location, represented by the radius of the ambiguity circle \cite{elbakly2016robust,elbakly2016cone}. The retrieved location label, its confidence, along with collected WiFi scan are stored at a temporary fingerprint  database for later processing.  This process is continuously performed in the background of the system operation through a crowdsensing manner, increasing the system coverage and keeping the fingerprint up-to-date.

 \begin{figure}[!t]
\centering
\includegraphics[width=0.4\textwidth]{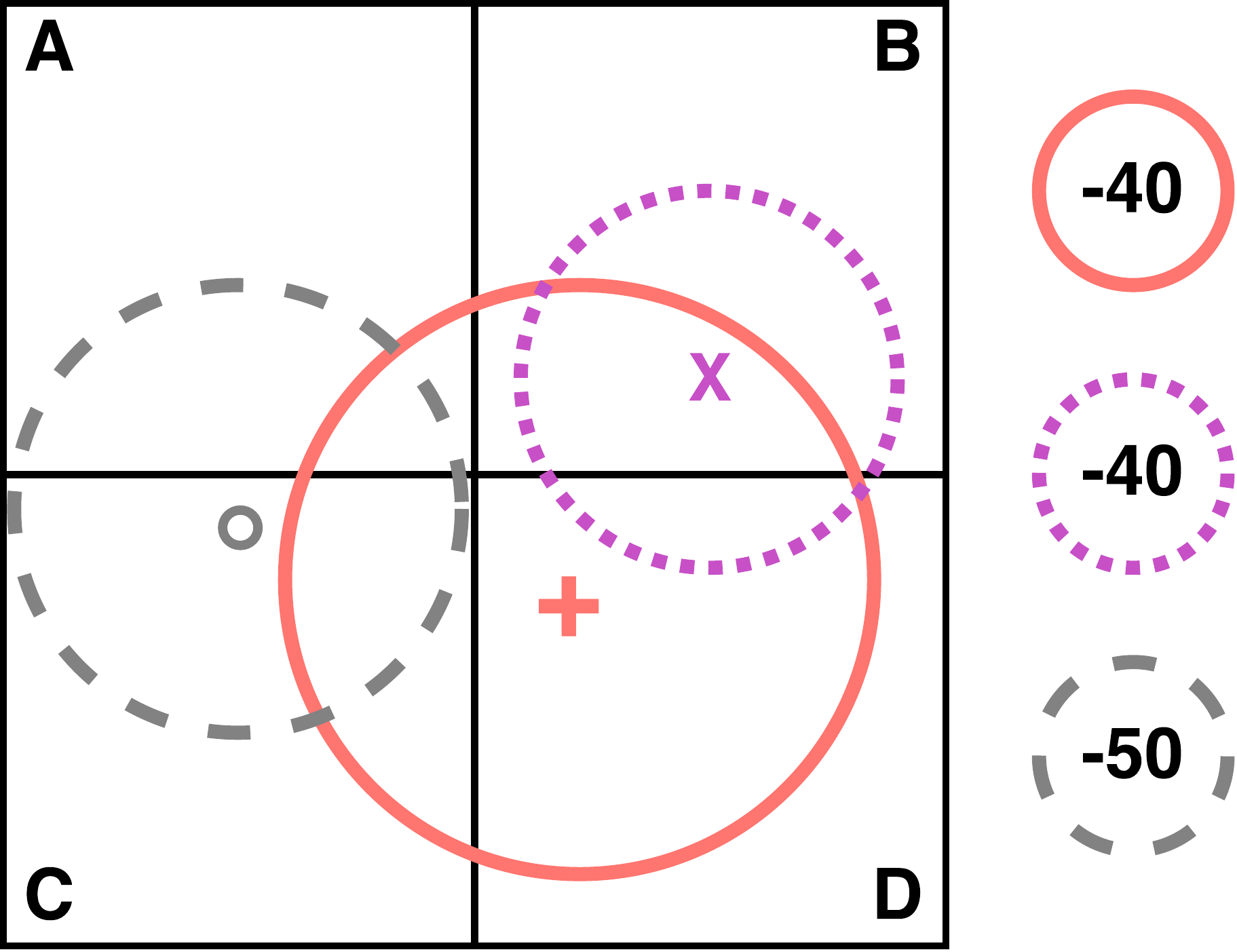}
\caption{Example showing the different techniques for assigning a WiFi scan to a grid cell based on the estimated BLE ground-truth location (center) and its estimated confidence (circle). The numbers represent the heard RSS in each scan.} \label{fig:samples}
\end{figure}

\begin{figure}[!t]
        \centering
        \begin{subfigure}[b]{0.25\textwidth}
                \includegraphics[width=1\textwidth]{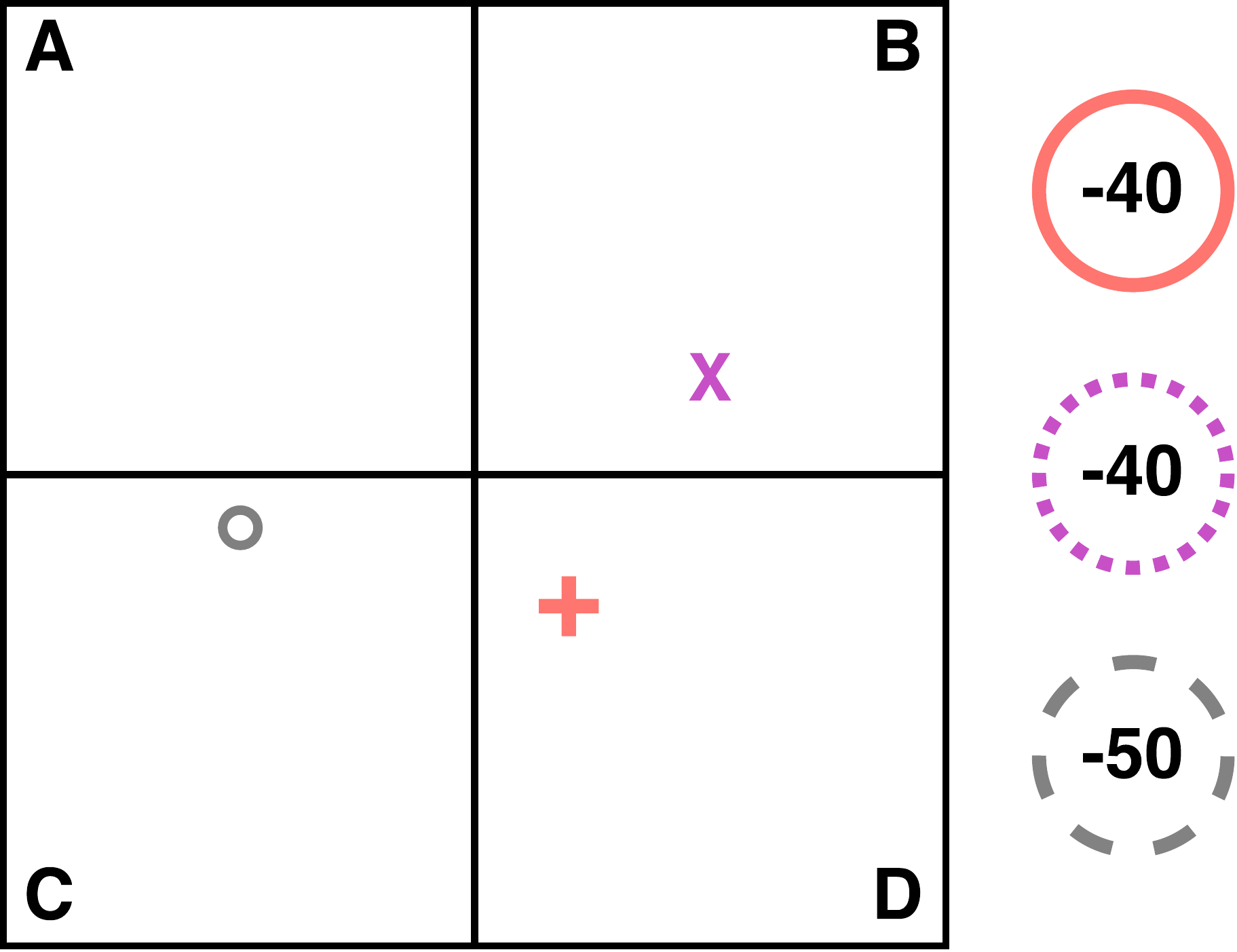}
           \caption{ground-truth locations for all scans.}
           \label{locationsch}
        \end{subfigure}%
        \begin{subfigure}[b]{0.25\textwidth}
                \includegraphics[width=\textwidth]{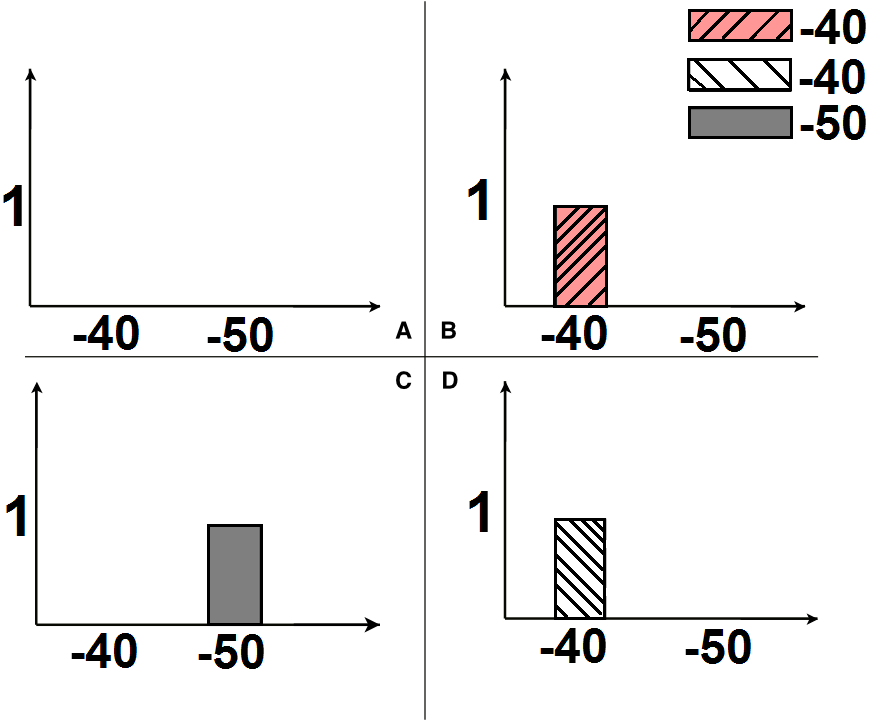}
              \caption{Histogram.}
            \label{locationhist}
        \end{subfigure}
        \caption{WiFi samples assignment to cells generated by  the location only approach. Subfigure a shows that the confidence circles are ignored and the estimated location is deemed as the  ground-truth location.  In this case, each WiFi sample is assigned to the cell where  the estimated location lies in as shown in subfigure b.}
             \label{fig:loc}
\end{figure}
\begin{figure}[!t]
        \centering
        \begin{subfigure}[b]{0.25\textwidth}
                \includegraphics[width=1\textwidth]{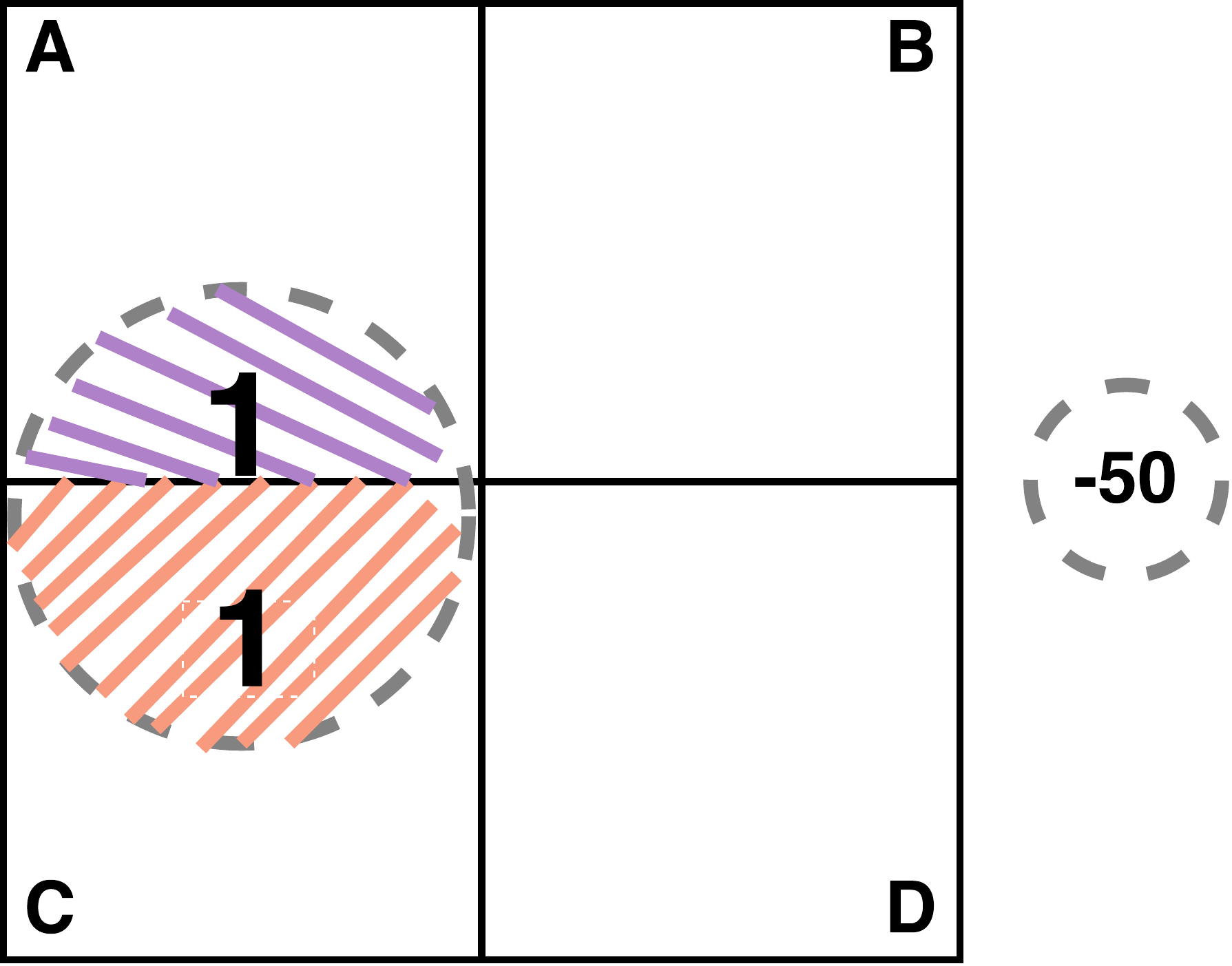}
           \caption{ground-truth location and confidence for the RSS=-50 scan.}
           \label{uconfidsch}
        \end{subfigure}%
        \begin{subfigure}[b]{0.25\textwidth}
                \includegraphics[width=\textwidth]{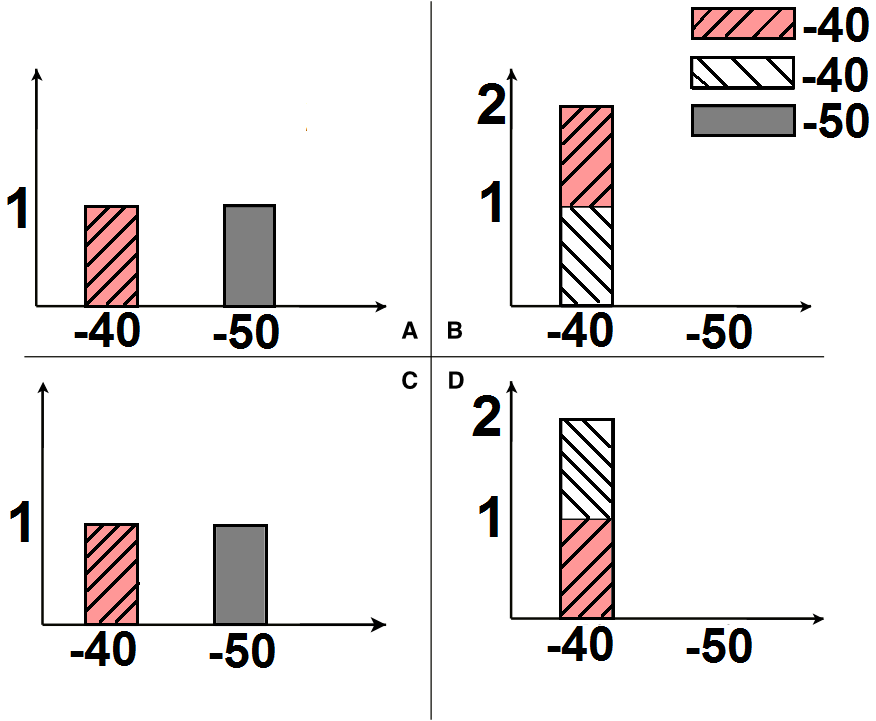}
              \caption{Histogram.}
            \label{uconfidhist}
        \end{subfigure}
        \caption{WiFi samples assignment to cells  generated by  the unweighted confidence approach. Subfigure a shows that  the  ground-truth location for RSS= -50 will be assumed to be in the two cells intersecting with the confidence circle.  In this case, the WiFi sample is assigned with an equal weight to all cells  where  the estimated location lies  in as shown in subfigure b. }
             \label{fig:uconf}
\end{figure}

\begin{figure}[!t]
        \centering
        \begin{subfigure}[b]{0.25\textwidth}
                \includegraphics[width=1\textwidth] {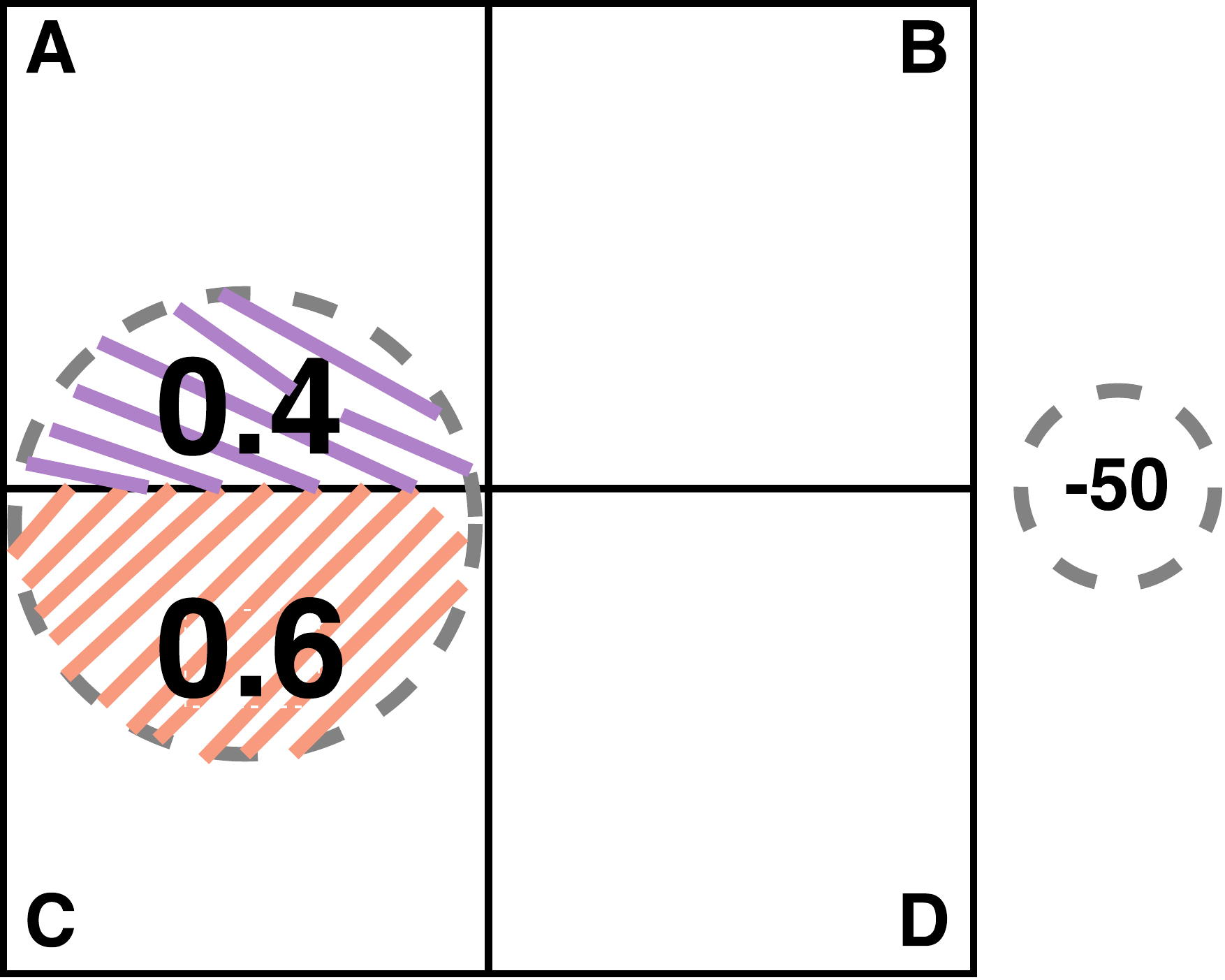}
           \caption{Ground-truth location and confidence for the RSS=-50 scan.}
           \label{wconfidsch}
        \end{subfigure}%
        \begin{subfigure}[b]{0.25\textwidth}
                \includegraphics[width=\textwidth]{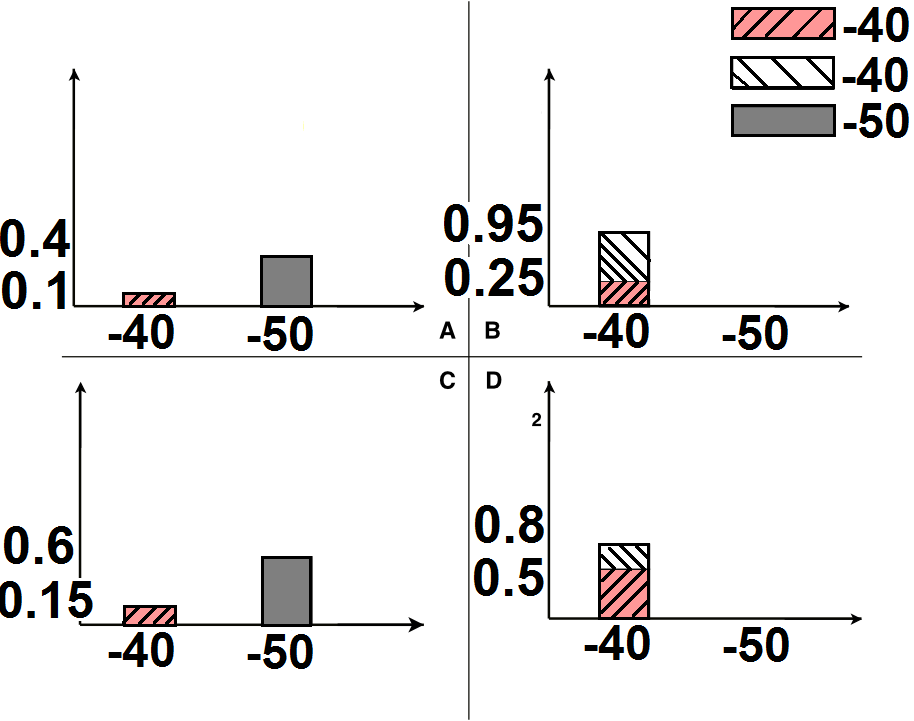}
              \caption{Histogram.}
            \label{wconfidhist}
        \end{subfigure}
        \caption{WiFi samples assignment to cells  generated by  the weighted confidence approach. Subfigure a shows that  the probability that a ground-truth location lies in a given cell depends on the intersection area of this cell and the confidence circle of that location. In this case, the WiFi sample is assigned to all cells  where  the estimated location lies in with different probability based on intersection area as shown in subfigure b. }
             \label{fig:wconf}
\end{figure}

\subsubsection{Handling ground-truth location noise} 
\label{sec:assignment}
This module is responsible for handling the inherent noise in the ground-truth BLE location. The input of this module is location-tagged WiFi scans and the estimated confidence of the BLE location. The goal is to determine the grid  cell(s) to assign these scans to. We experimented with  three different techniques  that  either  leverage the estimated location only,  fuse the location and the unweighted confidence, or fuse the location and the weighted confidence. 

   Without loss of generality, and for ease of explanation of these techniques and their differences, we present a simple deployment scenario where only one AP is  installed in the area of interest. Figure~\ref{fig:samples}   depicts   a part of the area of interest  divided into four grid cells A,B,C and D.  Assume that  the phone hears three  successive WiFi scans with RSS -40, -40 and -50 from the installed single AP with the estimated location labels and their confidence (the center of each circle is the estimated location and the circle represents the confidence in this location label). 
Now,  we illustrate  the difference among the three techniques by  highlighting  how they  assign  those three WiFi  scans to  the different grid cells:
  
 \begin{enumerate}
 \itemsep0em
\item{\textbf{Using the estimated  location only}\\}
 This technique assumes that the estimated BLE location is the true location. Therefore, it ignores the confidence estimate and assigns the WiFi scan to the grid cell enclosing the estimated BLE location.  This is illustrated in Figure~\ref{locationsch},  where   the centroids of the circles  (estimated locations) are deemed as the absolute locations of the collected samples.   The outcome of  the  assignment  using this  technique on the example in Figure~\ref{fig:samples} is shown in the histogram in Figure~\ref{locationhist}.

\item \textbf{Using the unweighted  confidence\\}
This  technique assumes that the user should be located anywhere inside the confidence circle. Therefore, it assigns the WiFi scan to \textbf{\textit{all}} grid cells that intersect with the confidence circle.  
Figure~\ref{fig:uconf} shows the assignment results for the RSS=-50 scan, where the scan is assigned to cells A and C. 
 
\item{\textbf{Using weighted confidence}\\}
This technique extends the previous one by taking the intersection area between the confidence circle and grid cell into account. The larger this intersection area is, the higher probability that the true location lies inside the cell. The assumption here is that the user can be located equally probable anyway inside the confidence circle. 
 \end{enumerate}

More formally, assuming that the user true location label is uniformly distributed inside the confidence circle, the grid cell weights can be calculated  as:
\begin{equation}
    w_i = \frac{1}{\pi c^2} \oint  \,dA_i = \frac{A_i}{\pi c^2}
\end{equation}

Where $c$ is the radius of the confidence circle and  $A_i$ is  the intersection area between cell $i$ and the confidence circle. 
Note that Monto Carlo Simulation can be used for more efficient estimate of the intersection area if needed.

Figure \ref{fig:wconf} shows the assignment result of the collected WiFi scans using this technique for the RSS=-50 sample. Note that the histogram is built using all the samples within each grid cell. Therefore, it will be normalized to a proper density function.

\subsubsection{Handling missing RSS}
\label{sec:missing_rss}
\begin{figure}[!t]
\centering
\includegraphics[width=0.4\textwidth]{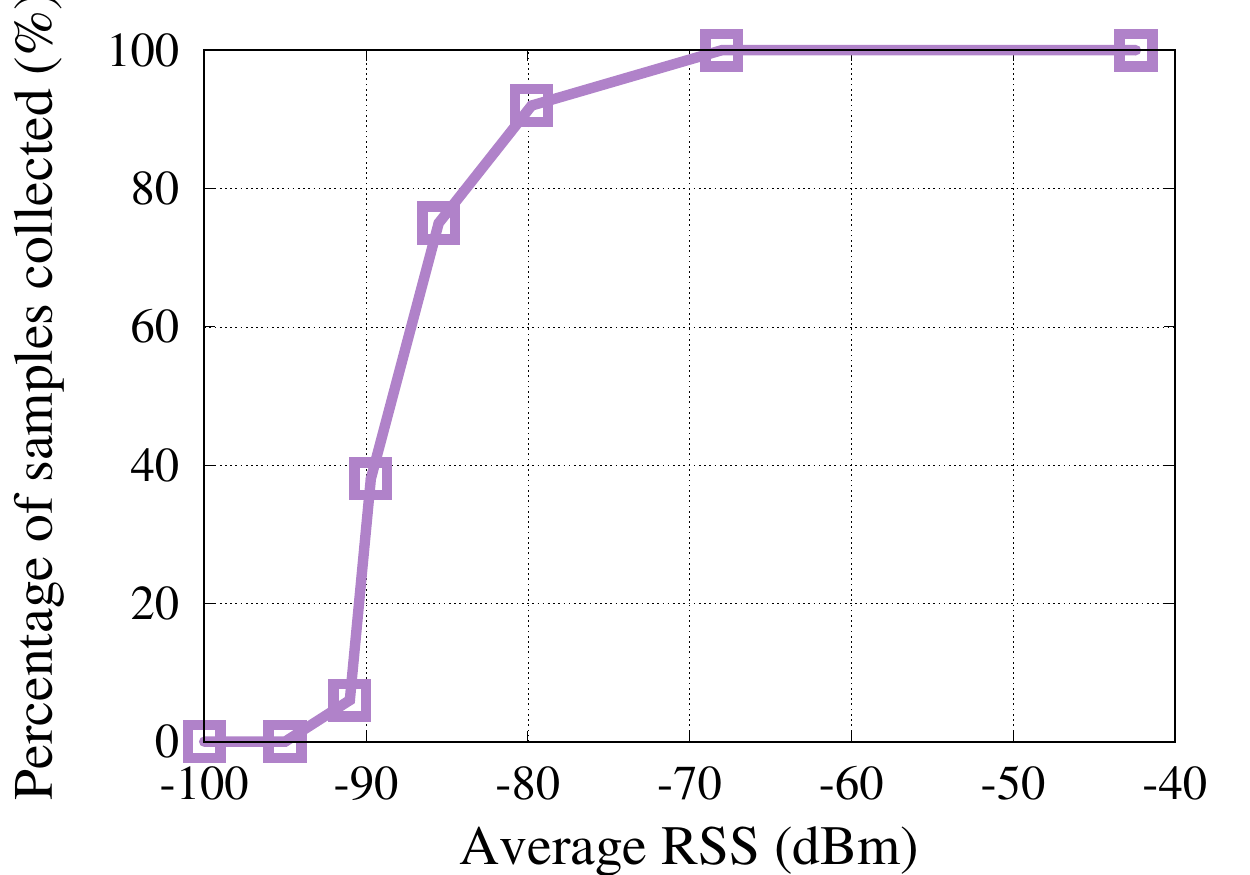}
\caption{Percentage of samples where the AP appeared versus its average RSS. The higher the RSS, the more samples expected from the AP. The sharp drop around RSS=-88 dBm is due to the receiver's sensitivity.}
 \label{missinghisto}
\end{figure}

\begin{figure}[!t]
\centering
\includegraphics[width=0.4\textwidth]{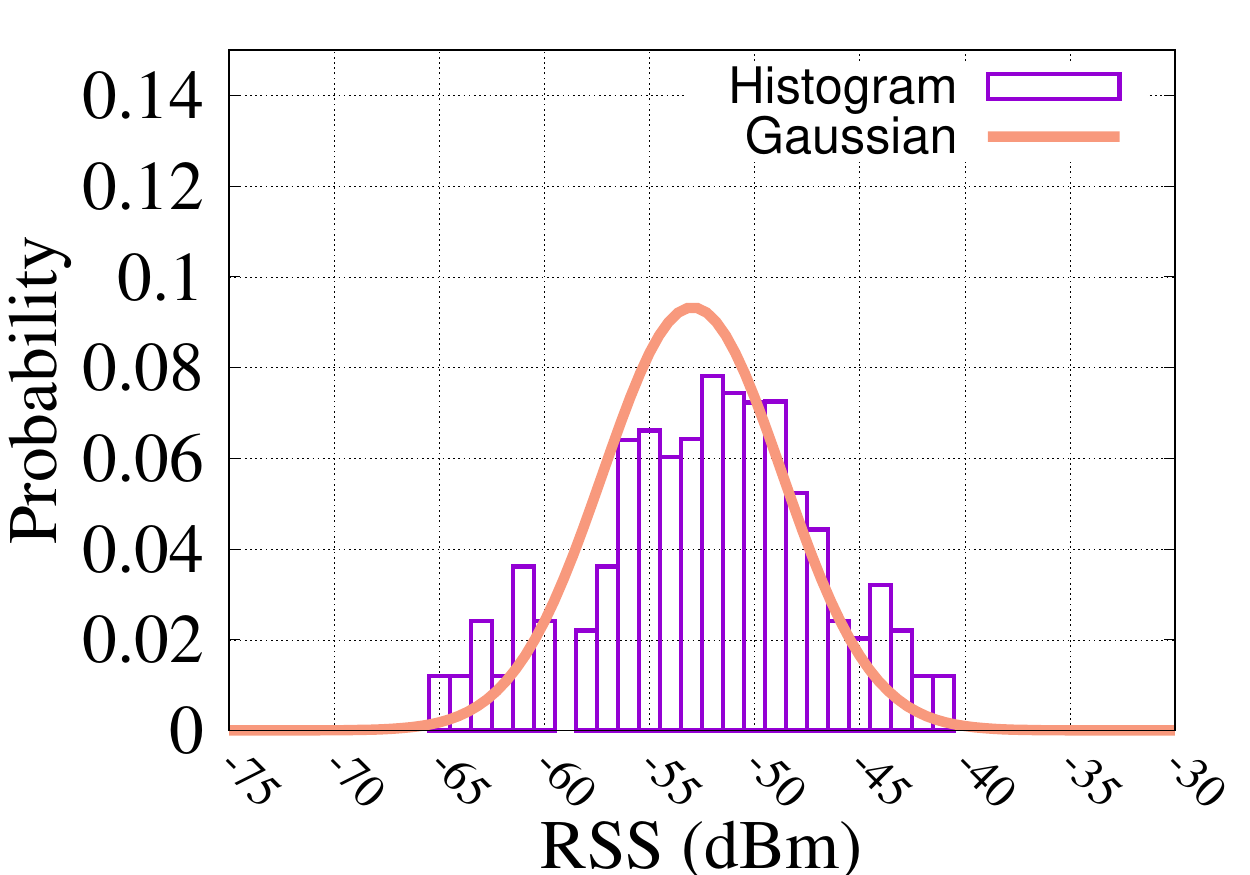}
\caption{Parametric and non-parametric representation of the RSS probability density function. The parametric distribution smooths the missing values, e.g. at RSS=-60 and is more space-efficient.}
 \label{fig:gauss_hist}
\end{figure}

After handling the noisy ground-truth location label problem, \sys{}  should  construct a   probabilistic WiFi fingerprint for each grid cell. However, due to the noisy wireless channel,  the number of APs in different scans may be different, especially for APs with a low RSS (Figure~\ref{missinghisto}). In addition, some RSS values may not be heard. To address these issues, we choose to use a parametric distribution  to estimate the signal-strength distribution. This leads to smoothing the distribution shape and avoids obtaining a zero probability for any signal strength value due to noise (Figure~\ref{fig:gauss_hist}). In addition, using a parametric distribution (i.e., Gaussian) is more memory-efficient than using a non-parametric distribution. 

More formally, \sys{} approximates the AP RSS histograms as a Gaussian distribution. Therefore,  the probability density of obtaining RSS $s_i$ from $AP_i$ at a specific grid cell ($g$) is given by:   
\begin{equation}\label{gaussDist}
P(s_i|g) = \frac{1}{{\sigma_i \sqrt {2\pi } }}e^{{{ - \left( {s_i - \mu_i } \right)^2 } \mathord{\left/
 {\vphantom {{ - \left( {x - \phi } \right)^2 } {2\sigma_i ^2 }}} \right.
 \kern-\nulldelimiterspace} {2\sigma_i ^2 }}}
\end{equation}

where $\mu_i$ is mean of all RSS values inside the grid cell $g$  
and $ \sigma_i$ is its standard deviation. 

\subsubsection{Handling devices heterogeneity}
 To handle the devices heterogeneity, one can build a fingerprint for each type of phones or a mapping function can be used to map the RSS values between the different types of cell phones  \cite{ibrahim2013enabling,zheng2008transferring}.  Nevertheless, the range of available user devices in the market, which keeps growing each day, makes this process not scalable and have a high overhead. Vaupel et al. \cite{vaupel2010wi} proposed a pre-calibration process for different devices to increase the performance of their localization system. However, pre-calibration  incurs an extra  overhead.
  \begin{figure}[!t]
\centering
\includegraphics[width=0.4\textwidth]{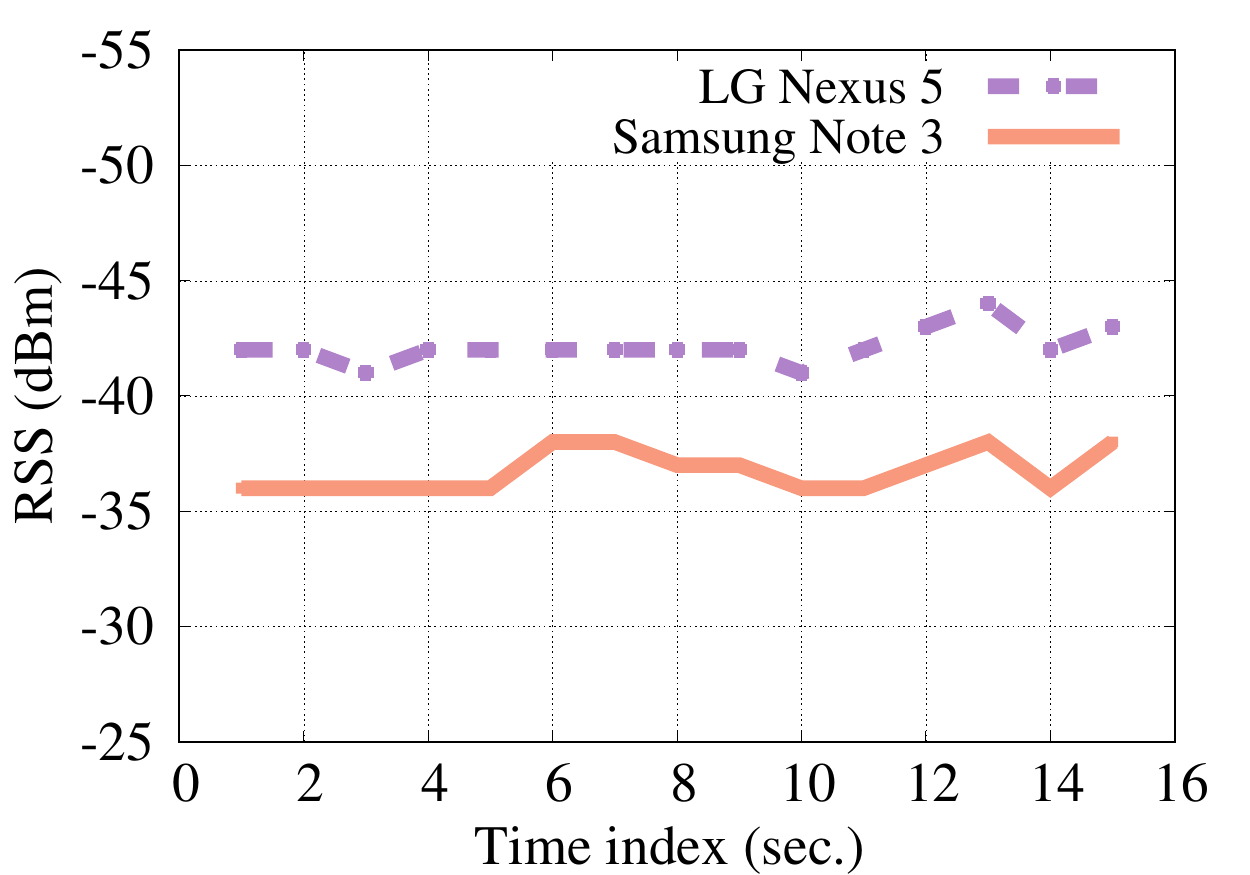}
 \caption{An example of WiFi RSS's collected by two different phone models at the same location from one AP.} 
 \label{shift}
\end{figure}

To tackle the device heterogeneity problem in an \textit{automatic manner without calibration}, we observe that there is an offset in signal strength collected by different phone models at the same location (Figure~\ref{shift}). This offset is almost constant across different APs and can be attributed to the difference in chip type, phone form factor, and/or antenna location and gain. Therefore, we estimate this offset as the average difference between the heard RSS and those stored in the fingerprint over all APs. More formally, the  \textit{common offset} $\lambda$   is estimated as:
 \begin{equation}
 \label{offset}
  \lambda = \frac{\sum_{i=1}^q (s_i-\mu_i)}{q}
\end{equation}
where $s_i$ is the RSS heard from  access point $AP_i$ by the device during the online tracking phase, $\mu_i$ is the mean RSS of the  $AP_i$  as recorded in the fingerprint database during the offline phase, and $q$ is  number of heard APs in this scan.
 
Therefore, we replace the random variable $s_i$ by the random variable $s_i- \lambda$. This new variable still follow a Gaussian distribution with mean $\mu_i- \lambda$ and variance  $\sigma_i^2 + \frac{\sum_{i=1}^q \sigma_i^2}{q^2}$.

Note that the same approach can be used to construct the signal strength distribution using heterogeneous phones during the offline phase. 
\subsection{Tracking Users- Online Phase}
In this phase, \sys{} works in two subsequent modules to enable continuous tracking of the user location as they move freely in the area of interest.  The first module, the \textbf{Discrete Location Estimator} module, determines the fingerprint cell that has the maximum probability given the received signal strength vector from the different access points. The second module, is the \textbf{Continuous Location Estimator} module, processes the discrete estimated user location returned by the previous module and returns a more accurate estimate of the user location in the continuous space. The rest of this section discusses the details of each module.
\subsubsection{Discrete-Space Estimator Module}
During the online phase, a user is stationed  at an unknown grid cell  $g\in \mathbb{G}$ hearing  a WiFi  scan with signal strength vector  $s = (s_1, ..., s_q)$ from the nearby $q$  APs, where  $s_i$ is the RSS measurement from $AP_i$. We want to find the grid cell $g^*$ that has the maximum probability given the received signal strength vector $s$. That is, we want to  find:
\begin{equation}
   g^*= \argmax\limits_g[P(g|s)]
\end{equation}
Using Bayes' theorem this can be rewritten as:
\begin{equation}
    g^*= \argmax\limits_g[P(g|s)] = \argmax\limits_g[\frac{P(s|g)P(g)}{P(s)}] 
    \label{eq:Bayes}
\end{equation}
Assuming all grid cells are equally probable\footnote{If the user profile over the different cells ($P(g)$) is known, it can be used directly in Equation~\ref{eq:Bayes}.}, this can be simplified as:
\begin{equation}
 g^*= \argmax\limits_g[P(g|s)] = \argmax\limits_g[P(s|g)] 
\end{equation}
The $P(s|g)$ term can be calculated using the Gaussian densities that have been constructed during the offline phase as: 
\begin{equation}
    P(s|g) = \prod_{i=1}^{q} P(s_i|g) 
\end{equation}

A representative location ($g_l$) for the most probable grid cell ($g^*$) is returned as the estimated location. This location can be simply the grid geometric center. However, we found that representing the cell by the center of mass of the points inside it gives a better estimate as quantified in Section~\ref{sec:eval}.

 \subsubsection{Continuous  Location Estimator Module}
 The previous module returns the representative location of the most probable grid cell as the estimated user location. However, if the user is moving normally in space, using this module only will lead to having the estimated location jump from one cell location to the next. To reduce this effect and enable tracking the user in the continuous space, this module aims to smooth the estimated discrete user location. To achieve  that, it uses two complementary approaches:

     \begin{enumerate}
 \itemsep0em
   \item \textbf{Center of mass of the estimated discrete locations (Spatial average):}
    Since the ``Discrete Space Estimator Module'' calculates the probability of each grid cell $g \in \mathbb{G}$, the first approach is to estimate the user location as the center of mass of all grid locations, taking the probability of the cell as its weight. More formally, let $P(g_l)$ be the probability of the representative  location of cell $ g \in \mathbb{G}$, the center of mass technique estimates the current user location $l$ as:
    \begin{equation}
      l=\frac{\sum\limits_{g \in \mathbb{G}} p(g_l) g_l} {\sum\limits_{g \in \mathbb{G}}p(g_l)}
   \end{equation}

  \item \textbf{Time averaging of location estimates (Temporal Average):}
This technique uses a time-average window to smooth the resulting location estimate. It calculates the average of  the last $k$ location estimates to obtain the final location estimate.  More formally, given a stream of location estimates  $ l_1, l_2,...,l_t$, the technique estimates the current location $ \bar{l_t}$ at time $ t$ as:
\begin{equation}
\bar {l}_{t}=\frac{1}{min(k,t)}\sum_{t-min(k,t)}^{t}l_{i}
  \end{equation}

\end{enumerate}

Note that both techniques are independent and can be applied together to further enhance accuracy.s
\section{Performance Evaluation} 
\label{sec:eval}
In this section,  we evaluate the performance  of  \sys{} in a typical indoor environment. We also describe the effect of different parameters on \sys{} performance.  Finally, we compare our system with the traditional manual fingerprint techniques.
\subsection{Data Collection}
\begin{figure}[!t]
\centering
\includegraphics[width=0.5\textwidth]{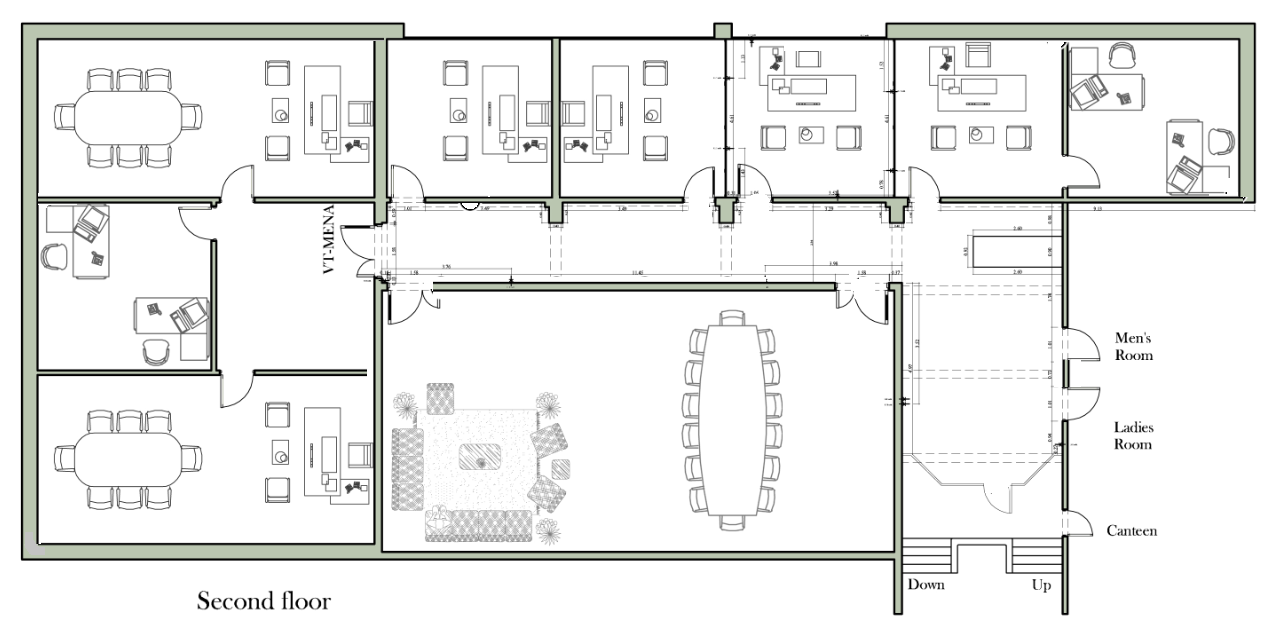}
 \caption{Environment testbed} \label{floor}
 \end{figure}
To collect the necessary data for evaluation, we deployed our system in a floor of our university campus building with a  $37m\times 17m$ area  containing offices, labs, meeting rooms as well as corridors (Figure \ref{floor}). The ceiling height is 3m.  We installed 20 iBeacons  at the same height of 2.5m uniformly  across the rooms  with an average density of  one beacon/33 $m^2$. We  use  the already installed  WiFi infrastructure  in the building, mainly four APs in addition to 12 APs overheard from other floors/buildings.     The data is collected by four participants  using different Android phones (e.g.,  LG Nexus 5, Samsung Galaxy Note 3,  Samsung Galaxy 4, Galaxy Tab, among other). 
 Two independent data sets are collected for constructing the fingerprint and evaluating the system. This captures the time-variant nature of the WiFi fingerprint as well as the heterogeneity of users and devices

We implemented a scanning program  using the Android SDK to simultaneously  scan APs and beacons. The program records the (MAC address, RSS,  timestamp) for each heard WiFi access point and the (UUID-major, minor, Tx power, Bluetooth address) for each BLE beacon. The scanning rate was set to one per second.    
Test points were collected on a uniform grid with a 1m spacing. 
\begin{table}[!t]
\centering
\scalebox{0.73}
{
\begin{tabular}{|l|l|l|}\hline
   Parameter & Range & Default value \\ \hline \hline
   
   Cell size ($G_S$) & 1 - 64 (m)&1 \\ \hline
    Number of averaged  locations ($k$)& 1 - 16& 10 \\ \hline
   Number of training samples ($N_s$)& 100-700& 700\\ \hline
   Cell assignment method& \pbox{10cm}{Location only, \\Unweighted confidence, \\Weighted confidence} &\pbox{10cm}{Weighted \\confidence} \\ \hline
 
  \hline\end{tabular}
  }
  \caption{Default parameters values.}
   \label{par}
\end{table}
\subsection{Effect of Changing \sys{} Parameters}
In this section, we study the  effect of the different parameters on the system performance including cell assignment method, the grid cell spacing, number of location  samples used in the training phase, ground-truth location accuracy, and the  number of most probable locations averaged to obtain the final location. Table \ref{par} shows the default parameters values used throughout the evaluation section.
\subsubsection{Effect of the cell assignment method}
\begin{figure*}[!ht]
\noindent\begin{minipage}[t]{0.31\linewidth}
\includegraphics[height=3.5cm, width=1\textwidth]{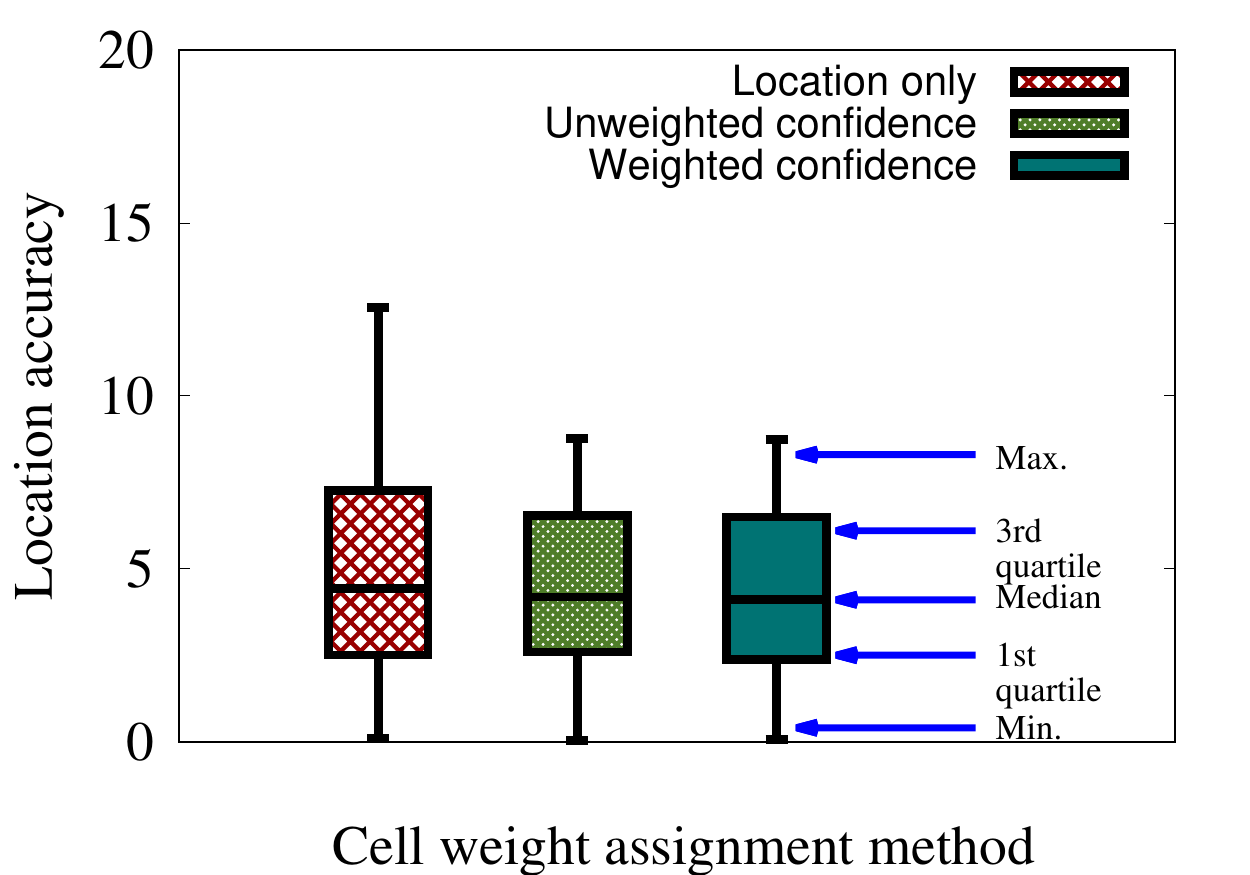}
 \caption{ Effect of the employed cell assignment method  on  the localization accuracy.}
  \label{boxplot}
\end{minipage}
\hfill
\noindent\begin{minipage}[t]{0.31\linewidth}
\includegraphics[height=3.5cm,width=1\linewidth]{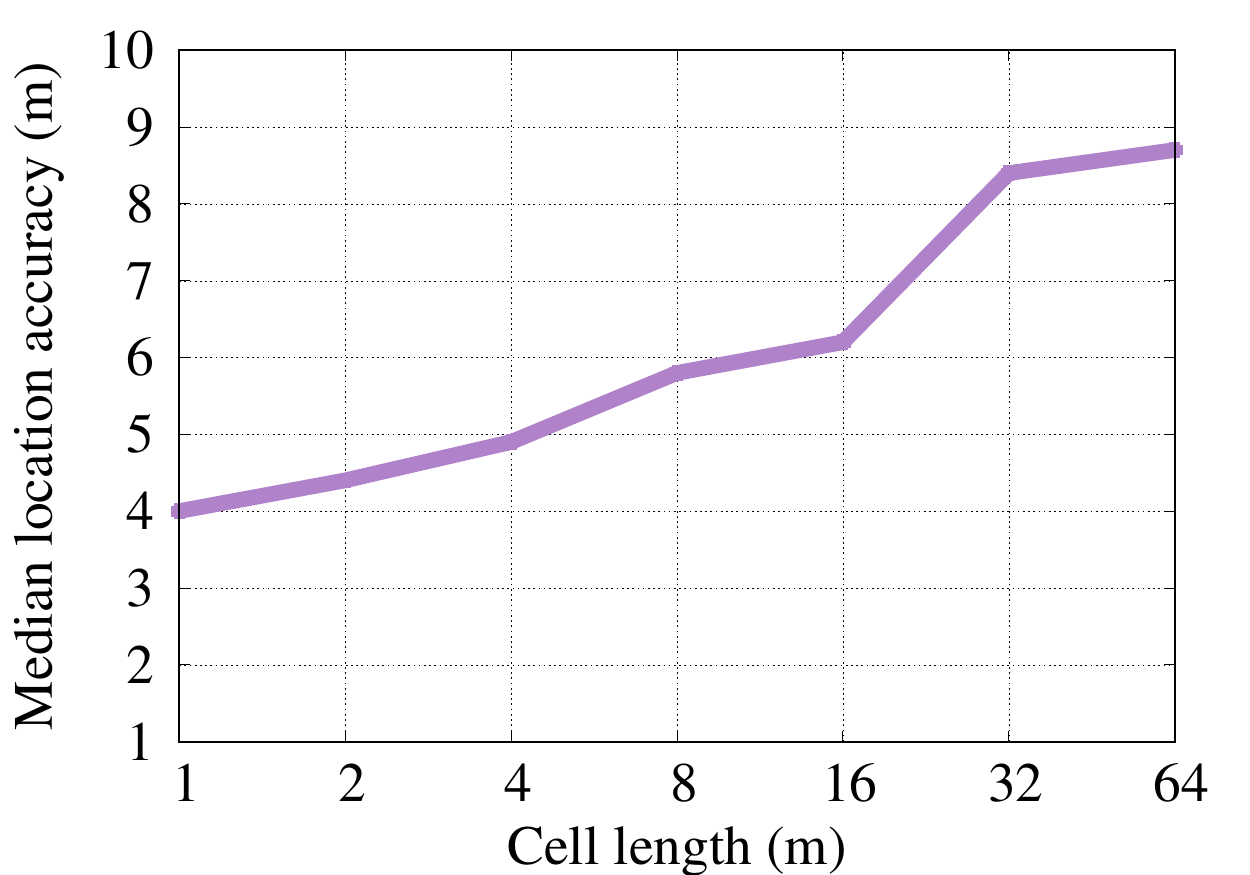}
\caption{Effect of the grid cell spacing on the median localization accuracy.}
\label{cellsize}
\end{minipage}
\hfill
\noindent\begin{minipage}[t]{0.3\linewidth}
	\includegraphics[height=3.5cm,width=1.1\linewidth]{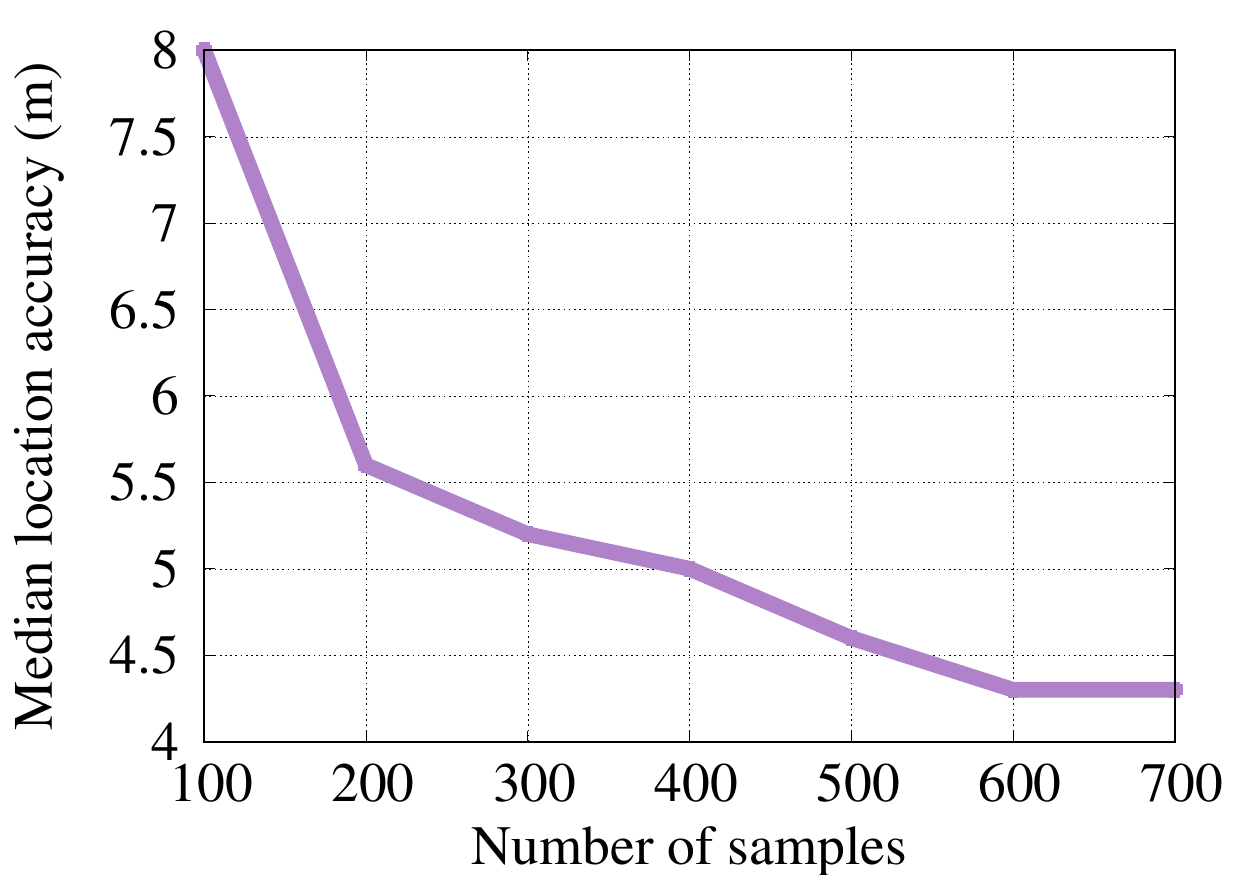}
	\caption{Effect of  number of training samples on the median localization accuracy.}
	\label{trainsamples}
\end{minipage}	
\end{figure*}

Figure \ref{boxplot} shows the box-plot for the localization accuracy  when using the three employed   cell  assignment methods described in Section~\ref{sec:assignment}.  The figure shows that the using the two confidence assignment methods lead to better accuracy as they handle the location noise better. 
 In addition, the ``unweighted confidence'' and the ``weighted confidence'' methods have comparable  localization accuracies  with a slight advantage  in favor of the weighted confidence approach. Therefore, the system designer can use the more computationally-efficient ``unweighted confidence'' technique.

\subsubsection{Effect of grid spacing ($G_S$)}
Figure \ref{cellsize} shows the effect of the grid spacing on the median localization  accuracy achieved by  \sys{}. The figure shows that, as expected, the localization accuracy increases with smaller cell lengths. This increase comes at the expense of a larger fingerprint and more time required to construct it due to the larger number of cells. However, this is performed only during the offline phase and; since \sys{} is a crowdsensing-based system;  is amortized over the number of system users.

\begin{figure*}[!ht]
\noindent\begin{minipage}[t]{0.3\linewidth}
\includegraphics[height=3.5cm,width=1.\linewidth]{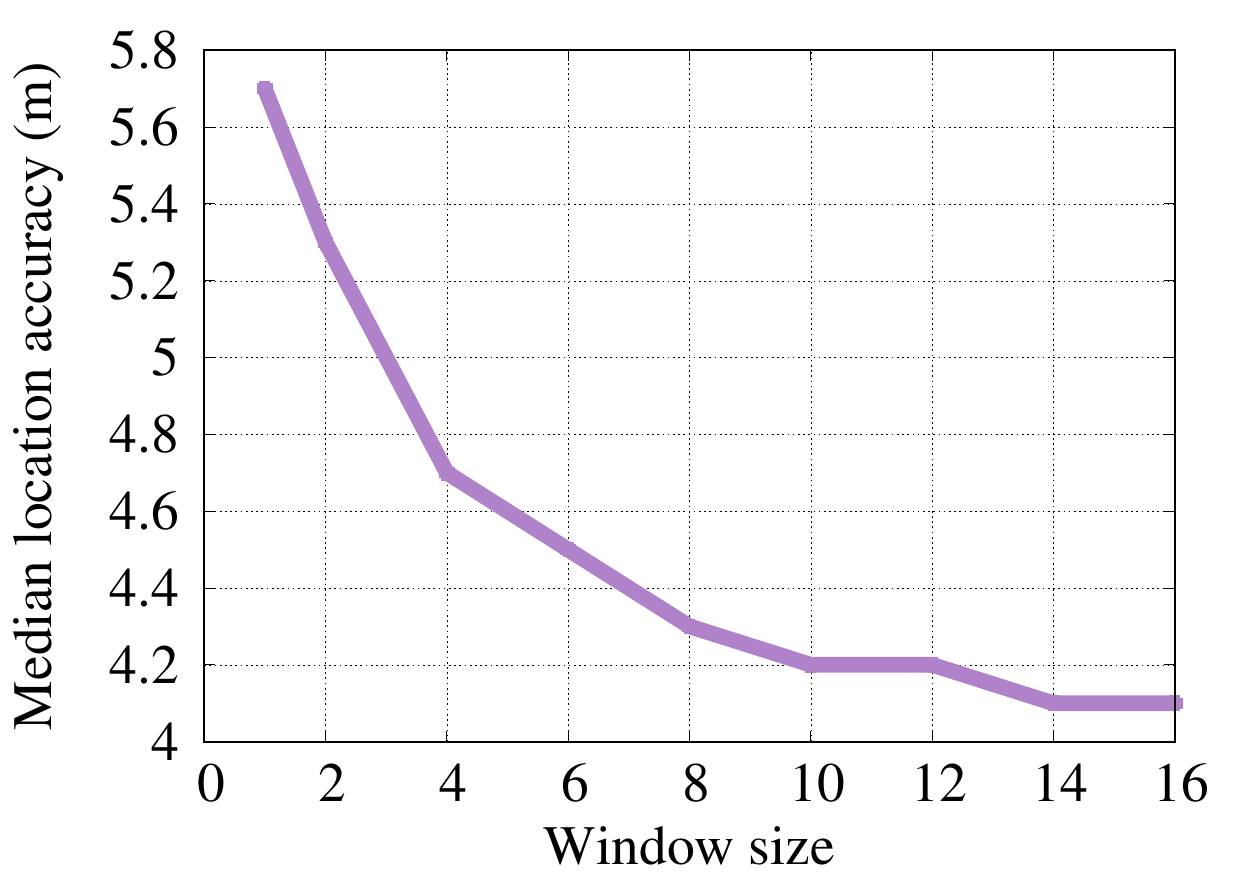}
\caption{Effect of the time-averaging window size on the median  localization accuracy.} 
\label{postprocessing}
\end{minipage} 
\hfill
\noindent\begin{minipage}[t]{0.3\linewidth}
\includegraphics[height=3.5cm,width=1.\linewidth]{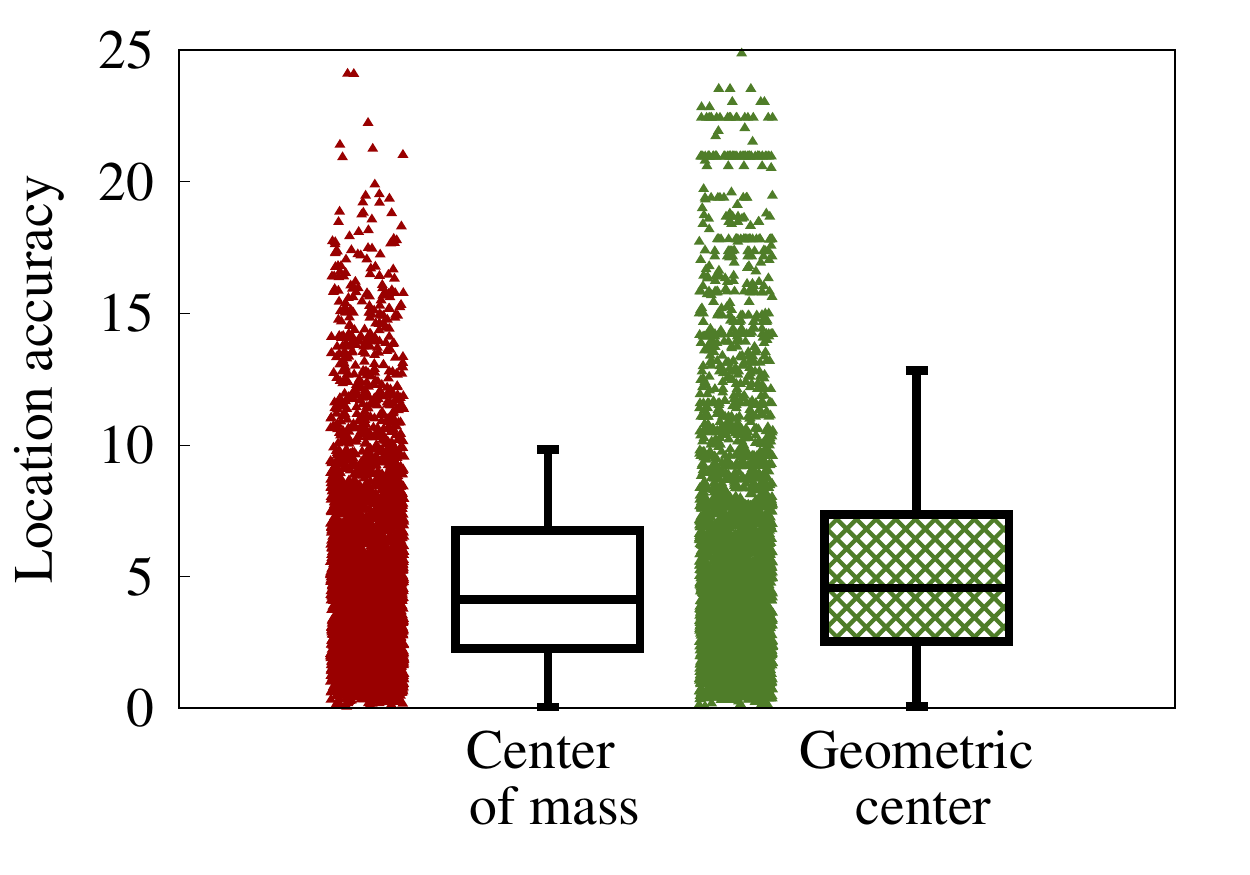}
 \caption{Effect of the cell location representation method on the localization accuracy.}
  \label{celllocation}
\end{minipage}
\hfill
\noindent\begin{minipage}[t]{0.31\linewidth}
	\includegraphics[height=3.5cm,width=1\linewidth]{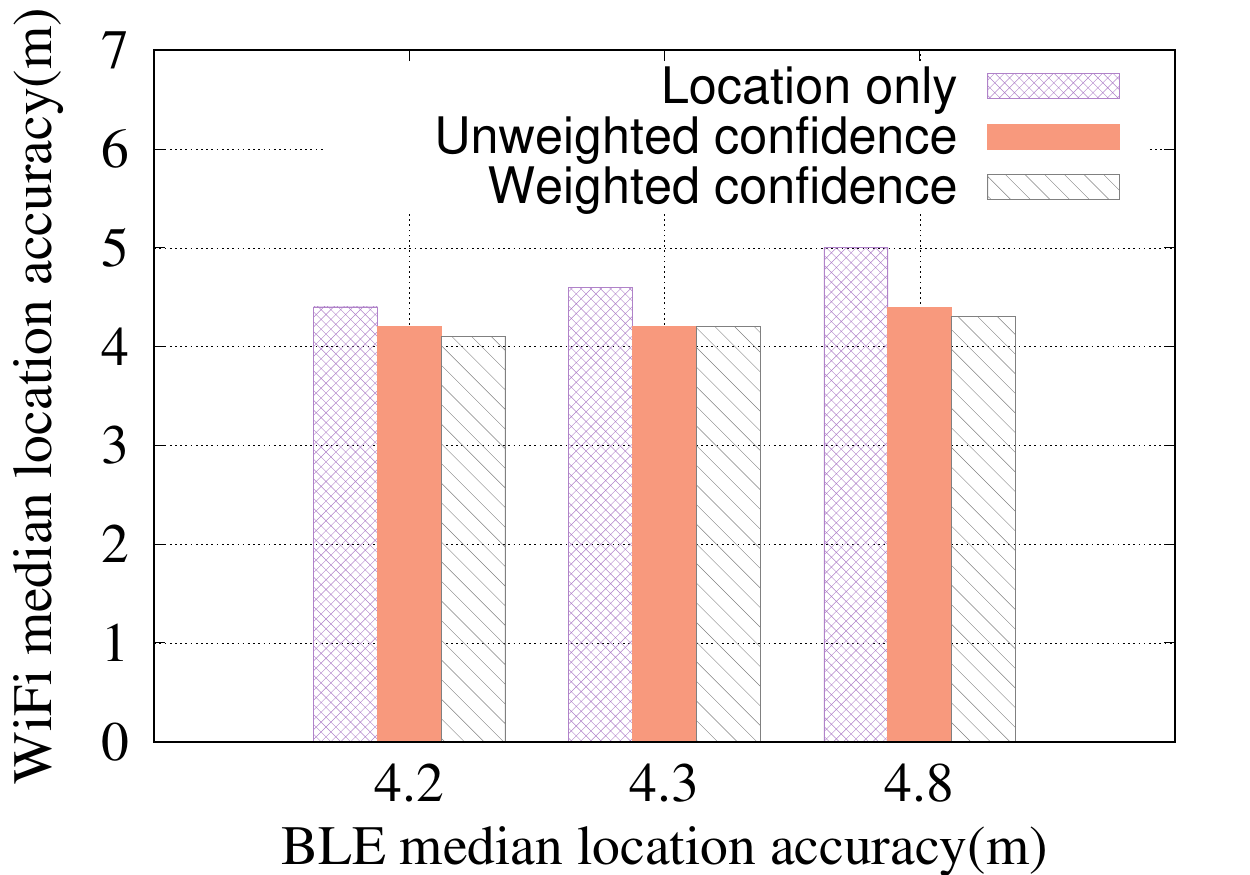}
	\caption{Effect of  the  ground-truth  accuracy on the WiFi localization accuracy.}
	\label{bleaccuracy}
\end{minipage}
 \end{figure*}
\subsubsection{Effect of the number of samples used for training ($N_s$)}
Figure~\ref{trainsamples} shows the effect of the  number of collected BLE ground-truth training samples from high-end phones on the median WiFi localization accuracy achieved by \sys{}.
 Evident from the figure, as the number of training samples increases, the  localization accuracy increases; till it saturates  when the number of samples reaches 600. This can be explained by noting that as the training samples density increases, the estimated histogram becomes  more accurate and representative of the fingerprint at that cell, leading to better accuracy. Note again that this is amortized over all system users.

\subsubsection{Effect of the number time-averaging window size($k$)}
Figure~\ref{postprocessing} shows the effect of the number of averaged locations in the continuous location estimator module on the system median accuracy.  
The figure shows that as more  location samples are aggregated to obtain the user location,  the localization accuracy  increases. 
However, using a large window size leads to  an increased latency in response to user movement.  Therefore, we have a trade-off between accuracy and latency of the location estimate that should be optimized based on the end user needs in a particular deployment. 

\subsubsection{Effect of  the cell  representative  location method}
As discussed in Section~\ref{sec:missing_rss}, the location that represents the grid cell may be the geometric center of the cell or the center of mass of all samples that are collected within that cell. Figure~\ref{celllocation} shows the box-plot of localization error for the two techniques. The figure confirms that the center of mass representation of the cell location  has a slight  improvement  over the  geometric cell center. This is because the former implicitly better captures the building geometry and where people move. 

\subsubsection{Heterogeneity Effect}
To demonstrate that \sys{} is robust to different heterogeneous phones due to its ``Devices Heterogeneity Handler'' module, we perform an experiment where different phones are used for training and testing. 
Specifically, we carried out experiments on two different phone models:  Samsung Galaxy Note 3 \textit{\textbf{tablet}} and Samsung S4 \textit{\textbf{smartphone}}. The two phones have a completely different form factor and  WiFi chips. 

Figure \ref{offset} shows the effect of employing the ``Devices Heterogeneity Handler'' module on the system accuracy when the Samsung Galaxy Note 3 tablet is used for training and the Samsun S4 smart phone is used for testing. The figure shows that the module does provide higher accuracy by 14\%.

We also study the effect of the module when using the same device (Samsung Galaxy Note 3)  for training and testing. Figure \ref{offset} shows even with the same model, the  module leads to a better localization accuracy.  This is due to reducing the effect of the noisy wireless channel. 

\subsubsection{Effect of the employed BLE system  accuracy}
To understand the   effect of the ground-truth locations from the  BLE indoor localization technology on \sys{}, we plot the WiFi localization accuracy at different accuracies of the ground-truth locations  provided by BLE-based technique (Figure \ref{bleaccuracy}). 
The figure shows that as BLE localization accuracy increases, the  WiFi localization accuracy increases too. The techniques that use the confidence estimate are more robust and less-sensitive to the changes in the accuracy of the employed BLE localization system. \textit\textbf{{Therefore, users with low-end phones with WiFi only can obtain comparable accuracy to those obtained using high-end phones with BLE.} }

\begin{figure}[!t]
\centering
\includegraphics[height=5cm, width=0.4\textwidth]{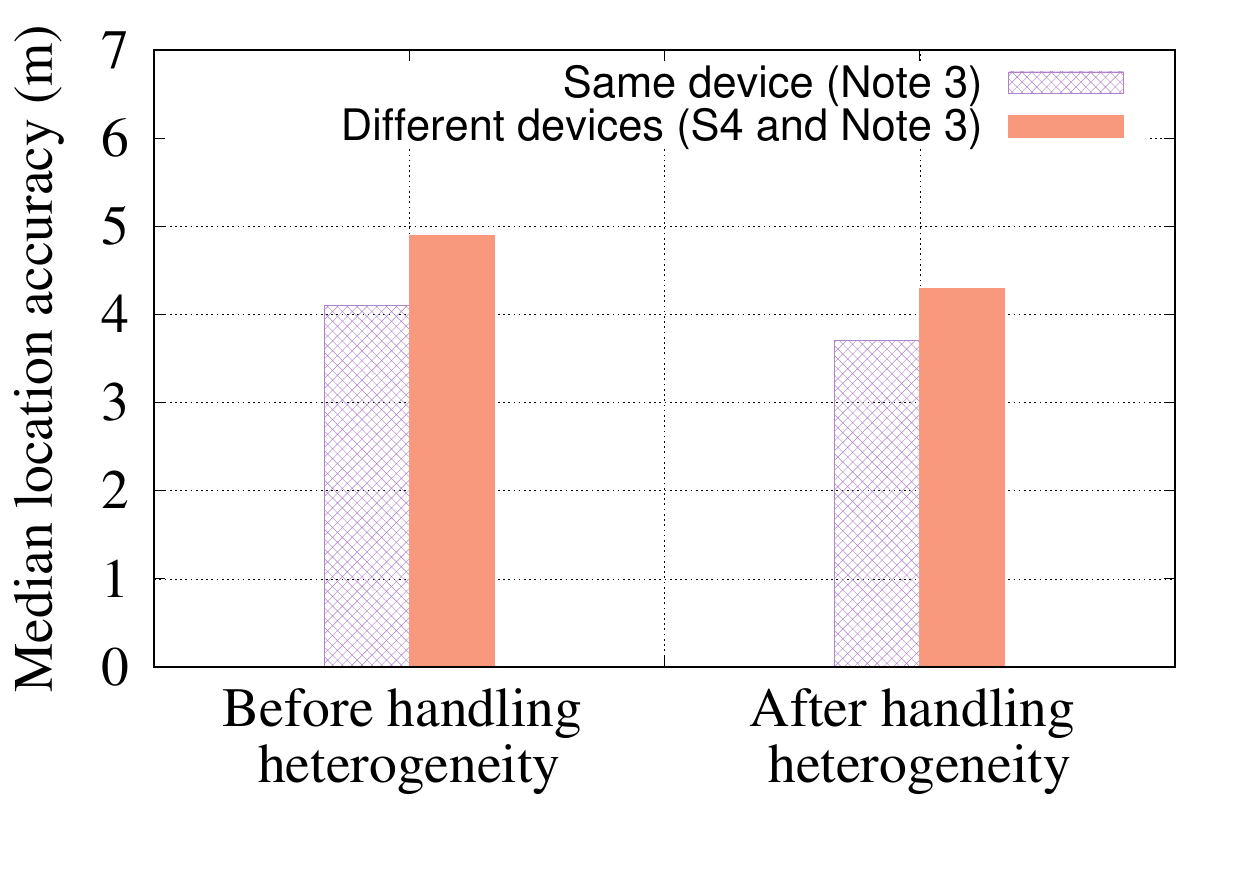}
 \caption{Effect of removing the common offset  using  the same  device (Note 3) and  different  devices  (training: Note 3, test : S4).}
  \label{offset}
 \end{figure}

\subsection{Comparison with Other Systems}
\begin{figure}[!t]
\centering
\includegraphics[width=0.45\textwidth]{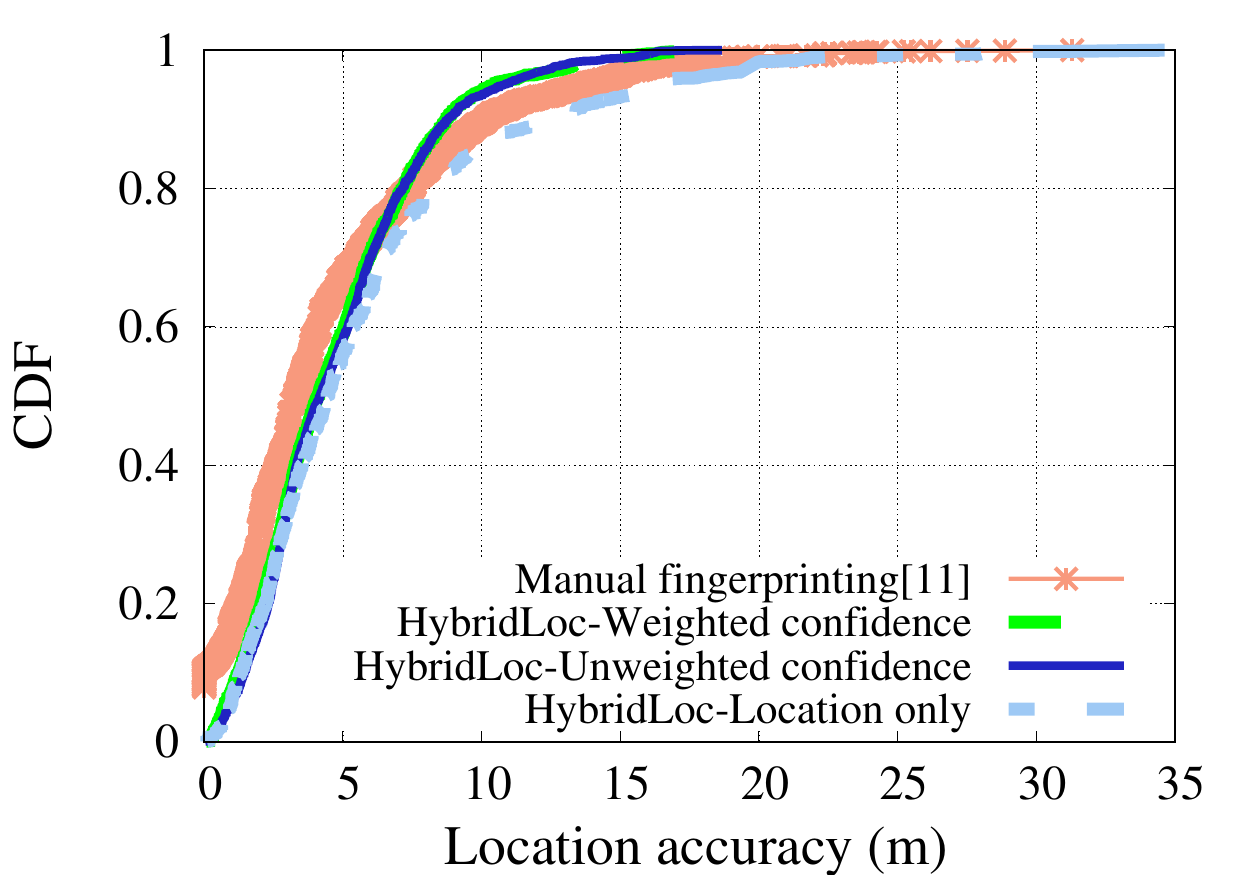}
\caption{CDF of localization error of different variants of \sys{} against a manual  fingerprinting  technique \cite{youssef2005horus}.}
\label{fig:cdf}
\end{figure}

In this section, we compare the location accuracy generated automatically by the three different methods  used in \sys{}  against a typical probabilistic  \textit{\textbf{manual}} fingerprinting technique (i.e., the Horus System \cite{youssef2005horus}). 
Figure \ref{fig:cdf} shows the CDF of localization accuracy for  these different algorithms. The figure  illustrates  that   the  confidence-based approach achieves approximately similar location accuracy as manual fingerprinting one as summarized in Table \ref{percentile}. This comes with no the costly or time-consuming calibration phase.

\begin{table}[!t]
\centering
\scalebox{0.73}{
\begin{tabular}{|l||l|l|l|l|}\hline
   Technique & \pbox{10cm}{$25^{th}$\\percentile }& \pbox{10cm}{$50^{th}$\\percentile}&\pbox{10cm}{$75^{th}$\\percentile}&\pbox{10cm}{$90^{th}$\\percentile}\\ \hline \hline
    \pbox{10cm}{Manual Fingerprinting \cite{youssef2005horus}}& 1.7&\textbf{3.3}&6.3&10.1\\ \hline\hline
       \pbox{10cm}{\sys{}-Weighted confidence}&2.4& \textbf{4.1}&6.4&8.7  \\ \hline
\pbox{10cm}{\sys{}-Unweighted confidence}& 2.5&\textbf{4.1}& 6.5&8.8\\ \hline       
   \pbox{10cm}{\sys{}-Location only}& 2.5& \textbf{4.4}&7.3&12.5 \\ \hline
  \hline
  \end{tabular}
  }
  
  \caption{summary of the location accuracy percentiles of different techniques}
   \label{percentile}
\end{table}

\section{Related Work}
\label{sec:related}
This section presents a brief background on the current  techniques for  WiFi-based indoor localization categorized into  fingerprinting-based, propagation models-based, and crowdsourcing-based techniques.
\subsection{Fingerprinting-based Techniques}
In these techniques, the radio map is built by  maintaining  the RSS signature of heard APs at different locations in a database during the offline phase.  During the tracking phase,  the set of  overheard  APs is matched against fingerprints  in the database  for the closest location in the RSS space to the unknown location.  This matching is done  using  either deterministic methods \cite{bahl2000radar,kontkanen2004topics}  or probabilistic methods \cite{youssef2005horus,youssef2008horus}. In the deterministic case,  the fingerprint  is represented by a scalar quantity, e.g. the average RSS of the heard APs in a certain  location. During the online phase, the  RSS vector collected while scanning the unknown location is matched  (based on some distance metric such as Euclidean, Manhattan, or Mahalanobis distance) against the fingerprints of all locations maintained in the radio map to find the nearest match \cite{honkavirta2009comparative}.  On the other hand, probabilistic techniques construct  signal strength histograms for the RSS received from each AP at each location in the area of interest.  During tracking, 
the fingerprint is used to calculate the probability of the RSS vector at the unknown location at each location stored in the radio map. The most probable location is used as the estimated location. Many variants  of probabilistic  WiFi fingerprinting based indoor localization have been proposed to improve the  performance. For instance, talking the high correlation between consecutive  signal strength samples from same APs  into
account can help to  achieve better accuracy \cite{youssef2004handling}. Another approach aims to detect small scale signal variations and  perturbs the signal strength
vector entries to overcome it, thus improving accuracy \cite{youssef2003small}.  Clustering locations that  share a common set of access points  will significantly reduce the computational overhead \cite{youssef2003wlan}. Finally, probabilistic techniques have been proven to be superior to deterministic techniques \cite{youssef2003optimality}. 

Although fingerprinting-based methods are relatively accurate, their deployment is impeded by the high cost of calibration phase and the user inconvenience. In addition, they need to handle the fingerprint differences between heterogeneous devices. 

\textit{\sys{}, in contrast, automatically constructs the fingerprint without requiring direct user participation nor calibration measurements. In addition, it has modules to address the devices heterogeneity issue.}
\subsection{Propagation Models-based  Techniques}
Modeling-based techniques try to capture the relation between signal strength and distance using a propagation model. Therefore,  they can automatically generate  the
fingerprint without expensive site surveying. For example, the Wall Attenuation Factor (WAF) model augments the free space path loss model  with the attenuation caused by walls to  handle the complex indoor propagation conditions \cite{bahl2000radar}. Specifically, the direct path between the transmitter and receiver is used to calculate the number of walls between them, where passing through each wall leads to signal attenuation by  a constant amount.

More sophisticated  propagation models use the ray shooting technique  to calculate the path of waves, augmenting reflections and absorption from walls and other objects into the model.   ARIDANE  \cite{ji2006ariadne} and Aroma \cite{eleryan2011synthetic} incorporate the 2D and 3D ray tracing techniques respectively to get better RSS estimation across  the designated area.  These models take
as an input the 3D floor plan of the area of interest, obtained
from CAD tools or generated automatically \cite{alzantot2012crowdinside,elhamshary2015semsense,elhamshary2014checkinside,elhamshary2016transitlabel,elhamshary2017fine}, and the location of APs. They then estimate
the 2D or 3D paths between the transmitter and receiver along with their
interactions with the materials in the environment\cite{el2010propagation} to compute the RSS values from all available APs at each reference
point on a grid.

Propagation models-based techniques, however, has lower accuracy than fingerprinting-based techniques, still require samples from the environment to calibrate the model, require the locations of the access points in the building,  require high computational requirements for ray tracing, and the model parameters still depend on the specific phone used for measurements.
\subsection{Crowdsourcing-based Techniques}
Another line of research on WiFi-based localization  tries  to reduce the calibration overhead by using crowdsourcing for the fingerprint construction  through explicit  \cite{park2010growing} or implicit  \cite{wang2012no,rai2012zee,abdelnasser2016semanticslam}  user feedback.  In \cite{park2010growing},  the user mobile  periodically gathers a fingerprint of nearby APs and  checks against  the signal strength map to determine the user's location. If it cannot do that with a certain accuracy, it prompts the user to indicate her current location on a displayed map. This on-the-fly surveying binds the fingerprint observed by the user to the relevant space. 

Although this reduces the  deployment burden of the system, prompting the user frequently  to gain more coverage/accuracy is  inconvenient.

To  avoid prompting the user,   Unloc \cite{wang2012no,abdelnasser2016semanticslam} and Zee \cite{rai2012zee} both leverage inertial sensors to get a rough estimate of the user location through dead-reckoning and associate a fingerprint with it.  To reset the  accumulated error, Zee performs map matching with the floorplan while Unloc leverages unique anchors in the environment, detected based on the multi-modal sensors signature. However, inertial sensors in commodity cell phones are noisy, reducing localization accuracy. In addition, their accuracy depends on the phone holding position and orientation \cite{mohssen2014s,mohssen2017humaine}, which is still an active area of research.  

\textit{\sys{}, on the other hand, does not depend on inertial sensors and provides accuracy comparable to manual fingerprinting techniques. In addition, it handles heterogeneous devices naturally.}

\section{Conclusion}
\label{sec:conc}
We presented \sys{}: a hybrid, accurate,  and low-overhead indoor localization system that works with heterogeneous phones. \sys{} leverages users with high-end phones in a crowdsourcing manner to automatically build the WiFi fingerprint. As part of \sys{}, it handles different practical deployment issues  such as handling the inherent error in the  BLE-based ground-truth location, missing RSS samples, and devices heterogeneity. 

Evaluation of \sys{} in a typical building using different Android phones shows that it can achieve a median distance error of 4.1m using WiFi only, which is comparable to this obtained using manual fingerprinting as well as high-end phones with BLE chips. However, \sys{} avoids the intensive, tedious and time-consuming calibration phase.

\section{Acknowledgement}
This work is supported in part by a Google Research Award.

\bibliography{main.bbl}
\bibliographystyle{IEEEtran}

\end{document}